\documentclass[fleqn,oneside]{article}
\usepackage{graphicx}
\usepackage{amsfonts}
\newcommand{\field}[1]{\mathbb{#1}}
\newcommand{\bn}{$\boldmath$ \nabla $\unboldmath$}
\usepackage{amssymb}
\usepackage{latexsym}
\usepackage{mathrsfs}
\newtheorem{Def}{\sc Definition}

\newtheorem{T}{\bf Theorem}

\newtheorem{Lem}{\bf Lemma}
\begin{document}

\begin{center}
{\Large\bf General ${\mathcal{K}}=-1$ Friedman-Lema\^itre models and the averaging problem
in cosmology.}

\vspace{0.4cm}

{\large Martin Reiris}.\footnote{e-mail: reiris@math.mit.edu}\\

\vspace{0.1cm}

\textsc{Massachussetts Institute of Technology.}\\

\end{center}

\vspace{0.6cm}
\begin{center}
\begin{minipage}[c]{13cm}
{\small\noindent We introduce the notion of general ${\mathcal{K}}=-1$ Friedman-Lema\^itre (compact) cosmologies and the
notion of averaged evolution by means of an averaging map. We then analyze the Friedman-Lema\^itre equations and the role
of gravitational energy on the universe evolution. We distinguish two asymptotic behaviors: radiative and mass gap. We discuss the
averaging problem in cosmology for them through precise definitions. We then describe in quantitative detail the radiative case, 
stressing on precise estimations on the evolution of the gravitational energy and its effect in the universe's deceleration. Also in 
the radiative case we present a smoothing property which tells 
that the long time $H^{3}\times H^{2}$ stability of the flat ${\mathcal{K}}=-1$ FL models
implies $H^{i+1}\times H^{i}$ stability independently of how big the initial state was in $H^{i+1}\times H^{i}$, 
i.e. there is long time smoothing of the space time\footnote{The word smoothing here is
referred to the decay toward zero of the space time Bel-Robinson curvatures (and therefore of the derivatives), 
and not to a gain in Sobolev regularity 
as in usual PDE terminology.}. Finally we discuss the existence of initial ``big-bang" states of large 
gravitational energy, showing that there is no mathematical restriction to assume it to be low
at the beginning of time.} 
\end{minipage}
\end{center}

\vspace{0.5cm}

{\center \section{Introduction}}

An implicit assumption of the Friedman-Lema\^itre cosmologies as models of the actual universe is that, because matter distribution at large scales 
(visible or not) appears to be ``to a good extent" homogeneous and isotropic, the large scale evolution of the universe 
should be modeled as driven ``to a good extent" by an exactly homogeneous and isotropic material distribution. The assumption, now known as 
the {\it averaging problem in cosmology}, needs quantitative approval or disproval (see \cite{E-vE}). Phrasing the problem in a question one asks: is the 
large scale evolution affected by the 
small scale structure?. The reason of the difficulty lies evidently in the nonlinearity of the Einstein equation. An averaged source of 
matter doesn't give rise necessarily to the average of the original solution. We will discuss this and other issues from the perspective of 
{\it general cosmological models}, i.e. the study of arbitrary solutions of the Einstein equation in the Hubble gauge (constant mean
curvature (CMC) gauge) provided with a set of Friedman-Lema\^itre equations giving the cosmological interpretation to the framework.

The standard ${\mathcal{K}}=-1$ FL cosmology describes the universe history by the evolution of the energy and pressure densities 
of the different type of matter present. Starting from a ``big-bang" 
where the densities and the space-time curvature blow up, the universe evolution is described as eternally expanding, with decaying 
densities and space-time curvature at a particular pace according to their matter type. Such a description is analytically possible due to the homogeneity and 
isotropy of the space which reduce the 
Einstein equations into a set of ordinary differential equations, the so called Friedman-Lema\^itre equations. We will deal here
with compact cosmologies, i.e. space-times with compact Cauchy surfaces of hyperbolic type. When speaking about homogeneity and isotropy of a compact cosmology we will refer to those properties in the universal cover solution.  In its formal terms the 
geometric structure of the space-time is described by a metric of the form ${\bf g}=-d\tau^{2}+a^{2}(\tau)/V_{H}^{\frac{2}{3}}g_{H}$ on 
a four-manifold $\field{R}\times \Sigma$ where $\Sigma$ is a compact hyperbolic manifold i.e. a manifold admitting a metric of constant 
negative sectional curvature and where $V_{H}$ is the volume of $\Sigma$ endowed with the unique hyperbolic metric (the one with sectional 
curvature equal to negative one). If the densities of energy and pressure of the material fields are $\rho(\tau)$ and $p(\tau)$ the FL 
equations are\\

\begin{equation}
{\mathcal{H}}^{2}=\frac{8\pi G \rho}{3}-\frac{{\mathcal{K}}V_{H}^{\frac{2}{3}}}{a^{2}},
\end{equation}

\begin{equation}\label{fli2}
\frac{a''}{a}=-\frac{4\pi G(\rho+3p)}{3},
\end{equation}  

\noindent where ${\mathcal{H}}=a'/a$ is the Hubble parameter and $G$ is the gravitational constant. These equations must be complemented 
with an equation of state $p(\rho)$. An obvious observation about these models is that they do not have any pure gravitational
degree of freedom besides the gravitational field generated by the matter present. This fact is seen by 
making $\rho=p=0$ and observing that in that case the solutions are flat. We call these flat solutions {\it flat cones} as they
can be obtained as quotients of a future light cone in Minkowski space-time. For non homogeneous and isotropic solutions there is no 
way to define which part of the gravitational field is generated
and which part is free, as those properties (if anything) would be potentially defined only in special solutions or in 
asymptotic regimes. There are simply two fields interacting, gravitation and matter. In this sense, the 
gravitational field adds a new degree of freedom to general cosmological models which needs to be quantitatively described. 

We have found that a satisfactory way to analyze arbitrary solutions to the Einstein equations on the light of the questions raised by
cosmology and those raised by the FL models themselves, is to introduce the notion of {\it general cosmological model: an arbitrary
solution to the Einstein equations in the Hubble gauge, provided with a set of Friedman-Lema\^itre equations giving its interpretative
cosmological meaning.} Unlike the FL models where the FL equations are enough to describe the evolution, in general cosmological models
one must rely on the full Einstein equations to predict the behavior of the terms involved in the general FL equations and therefore 
interpret the solutions in cosmological terms. One purpose of the article is to start a rigorous analysis of the general FL equations 
using the full Einstein equations.  

An arbitrary solution ${\bf g}$ on a space-time manifold $\field{R}\times \Sigma$ where $\Sigma$ is a compact hyperbolic manifold, 
is in the Hubble gauge if the mean extrinsic curvature $k$ of the equal time Cauchy surfaces is constant. The foliation $\field{R}\times
\Sigma$ is called the CMC foliation. It is well know to be unique, intrinsically defined, an with the mean curvature $k$ varying monotonically on it,
in particular $k$ or ${\mathcal{H}}$ (as we will se the Hubble parameter is ${\mathcal{H}}=\frac{-k}{3}$) can be taken as a time variable.
It is important to remark that unlike other gauges, the Hubble gauge is intrinsic, i.e. it is implicitly given by the solution. Let's write
the metric as

\begin{equation}
{\bf g}=-(N)^{2}dk^{2}+X^{*}\times dk+dk\otimes X^{*}+g,
\end{equation}

\noindent where $N$ is the {\it lapse function}, $X$ the {\it shift vector} and $g$ is a spatial three-metric on $\Sigma$. To write the
general FL equations one defines the radius $a(k)$ at the time $k$ as $a(k)=V(k)^{\frac{1}{3}}$ and the proper time $\tau(k)$ at the time $k$ through 
(see \cite{B2} for a related approach)

\begin{equation}
\frac{d\tau}{d k}=\frac{\int{N}dv_{g}}{V}.
\end{equation} 
   
\noindent With these definitions the FL equations (deduced from the Einstein equations, see subsection \ref{subsec3.1}) are
   
\begin{enumerate}

\item \begin{equation}
{\rm First\ FL\ equation:\ }{\mathcal{H}}^{2}=-\frac{\int_{\Sigma}{\mathcal{N}}Rdv_{g}}{6V}+\frac{\int_{\Sigma}{\mathcal{N}}(16\pi G\rho 
+|\hat{K}|^{2})dv_{g}}{6V},
\end{equation}

\item \begin{equation}
{\rm Second\ FL\ equation:\ }\frac{a''}{a}=\frac{-\int_{\Sigma}{\mathcal{N}}(4\pi G(\rho+3p)+|\hat{K}|^{2})dv_{g}}{3V}.
\end{equation}

\end{enumerate}

\noindent Where ${\mathcal{N}}=\frac{N}{\bar{N}}$ (bar denotes volume-average) and has average equal to one. The derivatives, denoted with a prime
are proper time derivatives, i.e. $'=\frac{d}{d\tau}$. $\hat{K}$ is the traceless part of the second fundamental form $K$. 
Compared with the second FL equation \ref{fli2} in a perfect FL cosmology we observe the appearance of the weight term ${\mathcal{N}}$ 
which inexorably couples matter to gravitation and a purely gravitational term of 
$|\hat{K}|^{2}$ which is essential and represents the additional gravitational degree of freedom mentioned before. A particular solution 
is a FL model iff $\hat{K}=0$ and ${\mathcal{N}}=1$. 

On the light of general cosmological models, a fundamental question is to quantify the evolution of the different terms that appear in the FL equations. 
It is important to realize that the ultimate goal would be to understand the FL equations 
for solutions which are realistic at small scales, i.e. at the natural scale of the flow. This is a difficult problem, 
however we will argue that we can have an starting point if precise assumptions are made. Namely, in subsection \ref{subsec3.5} we will
introduce {\it assumption} (C), a precise quantitative hypothesis on the behavior of arbitrary solutions at late times, from which we will
make explicit estimations of the different terms involved in the FL equations. Assumption (C) is a close relative of the {\it weak cosmic censorship
conjecture} of Penrose, conjecture stated in an asymptotically flat context. In rough terms, assumption (C) precisely describes a family
of solutions and divides it in two classes: {\it radiation} and {\it mass gap}. A radiative solution is an ideal solution in which no sort
of compact object emerges along evolution, i.e. universes filled only with radiation. We will study this case in detail, although only
for gravitational radiation. The technique may be applied to other radiative contexts as well. In this case the gravitational field
can safely be isolated from the rest, and one can safely interpret $\frac{|\hat{K}|^{2}}{16 \pi G}=\rho_{G}=p_{G}$ as the effective
energy and pressure densities of gravitational radiation. These densities are quantitatively studied along with the decay of ${\mathcal{N}}$ to
one. The estimates are given in Theorem 1 (see statement below) which in addition give estimates on the Bel-Robinson energies $Q_{i}$. Altogether
Theorem 1 provides a detailed structure of the radiative solutions and is one of the main result of this article. 

\begin{T} {\rm (Expansive smoothing and energy estimates).}\label{Smooth} Let $\Sigma$ be a compact and rigid hyperbolic manifold. 
There is an $\epsilon>0$ such that the Einstein  CMC flow of a cosmologically scaled initial state (i.e. with ${\mathcal{H}}=1$) $(g,K)$ 
with ${\mathcal{V}}-{\mathcal{V}}_{inf}\leq \epsilon$
and $\tilde{\mathcal{E}}_{1}\leq \epsilon$ has the following long time properties (take $t=\frac{1}{\mathcal{H}}$):

\begin{enumerate}

\item The limit $lim_{t\rightarrow \infty} t^{3}Q_{0}$ is finite and greater than zero.

\item there are $n_{i}\geq 0$ such that $\lim_{t\rightarrow \infty} \frac{t^{2i+3}}{(\ln t)^{n_{i}}}Q_{i}\leq\infty$ for $i\geq 1$.

\item for given $\gamma>0$, $\int_{t}^{\infty}\frac{\int_{\Sigma}|\hat{K}|^{2}dv_{g}}{u}du\geq Ct^{-(2+\gamma)}$.

\item $|\hat{K}|^{2}\leq Ct^{-4}$ pointwise (not volume averaged).

\end{enumerate}

\noindent In particular the cosmologically scaled flow of a $H^{i}\times H^{i-1}$ state (for any $i\geq 1$) as in the
hypothesis above converges in $H^{i}\times H^{i-1}$ to the canonical flat cone state $(g,K)=(g_{H},-g_{H})$.

\end{T} 

\noindent Theorem 1 is in PDE terminology a small data statement. The small data condition is stated as saying that the 
reduced volume ${\mathcal{V}}={\mathcal{H}}^{3}V$ is $\epsilon$-close to its infimum and the first Bel-Robinson energy ${\mathcal{E}}_{1}$
$\epsilon$-close to zero. These two conditions can be seen to be equivalent \cite{R} to the statement that the initial data $(g,K)$ is 
close in the Sobolev space $H^{3}\times H^{2}$ to the flat cone state $(g_{H},-g_{H})$ where $g_{H}$ is the unique hyperbolic metric (up 
to diffeomorphism). A hyperbolic 
manifold is called rigid if it doesn't admit traceless Codazzy tensors (see \cite{A-M} for a discussion). 
The topological condition of rigidity is important to get the precise estimates above. It is 
possible to get estimates in the non rigid case but they are different, in particular those on the gravitational 
energy.  The importance of rigidity is that it allows the control of the $H^{2}$ norm of harmonic metrics with respect to the 
hyperbolic metric (spatial gauge) only by their Ricci tensor.

The estimates in theorem \ref{Smooth} are compatible with what one would expect is a radiative behavior. According to the standard  FL models 
an exact radiative behavior would imply a pointwise decay on the gravitational energy density of the form

\begin{equation}
\frac{|\hat{K}|^{2}}{16\pi G}\approx \frac{1}{t^{4}}
\end{equation}

\noindent The estimate in items $3$ and $4$ in theorem \ref{Smooth} says that in some averaged sense the global 
gravitational energy decays with a rate between the radiative $t^{-1}$ and the faster $t^{-2}$. It would 
be interesting to improve (if possible) the estimate from below in item $3$. 

In rough terms the {\it mass gap} solutions can be described as those for which after a sufficiently long time there appear a finite set 
of isolated stationary solutions separating from each other and with radiation in between. This qualitative description is made quantitative
in {\it assumption (C)}. We analyze the averaging problem for these {\it mass gap}-solutions. A convenient setup for the analysis is
to define the notion of averaged space, a Lorentzian manifold constructed out the averaged parameters $a(k)$ and $\tau(k)$ of the
original solution. The averaging problem can be stated as asking to which extent the averaged space is close to a FL model. A remarkable
consequence of applying assumption (C) is that the second FL equation is estimated as
  
\begin{equation}\label{fund}
\frac{a''}{a}=-\frac{4\pi G(\bar{\mathcal{M}}_{ADM}+\bar{\rho}+\bar{\rho}_{G}+3(\bar{p}+\bar{p}_{G}))}{3}+
O(t^{-(3+\epsilon)})
\end{equation} 

\noindent where $\bar{\mathcal{M}}_{ADM}$ is the volume average of the ADM masses of the emerging stationary solutions, and $\bar{\rho},\bar{\rho}_{G},\bar{p},\bar{p}_{G}$
are the volume averages of the densities of energy and pressure of material and gravitational radiation respectively, filling the space in between. 
However one must remark that despite all the satisfactory equation \ref{fund} may look, it is based on an idealized assumption and on its 
apriori estimates which so far need to be justified. Also the quantitative description they provide is only asymptotically in time, and not
throughout the full evolution. 

A quantity underlying all the averaging formalism is the so called {\it reduced volume} \cite{F-M1}, defined above as ${\mathcal{V}}={\mathcal{H}}^{3}V$.
It decreases monotonically, and is bounded below by the topological invariant $V_{H}$. It has been used in \cite{R} to show the long time
geometrization of the Einstein flow under curvature bounds. Here it is manifested throughout the article in different forms. Its monotonicity is
shown to be equivalent to the universes deceleration and is used to get the estimate in item $3$ in Theorem 1. We will introduce an
use an equivalent quantity that we will call the global CMC energy defined as

\begin{equation}
E_{CMC}=\frac{1}{4\pi G {\mathcal{H}}}({\mathcal{V}}-{\mathcal{V}}_{inf}).
\end{equation}  

\noindent Rather remarkably, the CMC energy is shown to express the full ADM energy of the time-asymptotic evolution only in terms of
the total volume, the Hubble parameter and the topological invariant ${\mathcal{V}}_{inf}$. 
    
The contents and sections are organized as follows. In section \ref{sec2} we introduce the main equations for
the Einstein-CMC flow as well as Bel-Robinson energies and their main formulae. In section \ref{sec3} we introduce 
the averaged cosmological parameters and the Friedman-Lema\^itre equations. The treatment has no restriction on the
sort of matter. We introduce the
Newtonian gravitational potential $\phi$, its Poisson-like equation and reformulate the FL equation with it in subsection \ref{subsec3.2}. 
As it turns out the Newtonian
potential is the main field to estimate when the purpose is to estimate the universe deceleration and the
Hubble parameter as a function of red shift $z$. In subsection \ref{subsec3.3} we introduce the CMC global energy and
relate it in subsection \ref{subsec3.4} with the ADM energy in the weak field limit, analysis extended in subsection \ref{subsec3.5} 
to arbitrary solutions under assumption (C). In subsection \ref{subsec3.6} we discuss the averaging problem on the light of assumption (C) for
the mass gap regime. We will use the CMC energy to estimate the 
gravitational energy in section \ref{sec4}. Also in section \ref{sec4} we prove the main estimates of theorem \ref{Smooth}. 
The technique may be
thought as estimating the gravitational field through a Taylor expansion (in time) of the zero-order 
Bel-Robinson tensor and is a natural extension of the analysis in (\cite{A-M}). In section \ref{sec5} we 
construct ``big-bang" states of high gravitational energy showing that there is no 
mathematical reason
to assume a low gravitational energy at the initial ``big-bang" state. The dynamics of those states even in 
short times is a completely open problem, in particular it is not know whether the initial rate of expansion with 
respect to proper time is of matter, radiation or of a type like non of both. In section \ref{sec7} we give an account of the main
points of the article.    

{\center \section{The CMC flow equations and the Bel-Robinson energies.}\label{sec2}}

{\center \subsection{The CMC flow.}\label{subsec2.1}}

\vspace{0.1cm} 
In this section we consider the formal setup of the Einstein CMC flow equations. A detailed account can be found
in \cite{R}. Consider $\Sigma$ a compact hyperbolic three-manifold. A cosmological solution to the Einstein equations 
with compact Cauchy surface $\Sigma$ is formally a Lorentz metric ${\bf g}$ on a four manifold of the 
form $I\times \Sigma$ (where $I$ is a interval) and where the equal time hypersufaces $\Sigma_{t}$ are space-like
i.e. the induced metric is Riemannian. If the mean extrinsic curvature ($k=tr_{g}K$) is constant on each slice of the foliation 
$\{\Sigma_{t}\}$
then we say that the cosmological solution is in the (temporal) CMC-gauge. When the spatial topology is a hyperbolic
manifold the mean curvature $k$ cannot be zero (due to the energy constraint and the fact that $\Sigma$ doesn't accept 
metrics of non-negative scalar curvature) and it can be proved to be strictly monotonic over a unique and 
connected interval.  
For a three manifold of hyperbolic type in particular it is conjectured that the CMC foliation has a range of $k$ equal to
$I=(-\infty,0)$, i.e. from a ``big-bang" when ${\mathcal{H}}\rightarrow \infty$ towards an infinitely expanding universe when 
${\mathcal{H}}\rightarrow 0$. Say $\partial_{t}=NT+X$ where $T$ is the unit normal to the slices  and $t=k$. Write the four metric as

\begin{equation}
{\bf g}=-N^{2}dt^{2}+X^{*}\otimes dt+dt\otimes X^{*}+g,
\end{equation}

\noindent where $g$ is the spatial three dimensional metric. $N$ is called the lapse and measures the
rate of proper time with $k$ (locally). $X$ is called the shift vector field and can be chosen freely but compatible
with the regularity. For a discussion of the initial value formulation in the CMC gauge we refer the reader to \cite{R}. 
We call the path $(g,N,X)(k)$ the CMC flow. A CMC state is a pair position-normal velocity $(g,K)$ (where $K$ is 
the second fundamental form and is equal to $K=-\frac{1}{2}{\mathcal{L}}_{T}g$) with $k=tr_{g}K$ constant. Thus the CMC flow gives rise 
to a flow of position and velocities $(g,K)(k)$. With this notation the Einstein equations

\begin{equation}
{\bf Ricc}-\frac{1}{2}{\bf R}{\bf g}=8\pi G {\bf T},
\end{equation}

\noindent can be seen as the CMC flow equations (taking $t=k$)

\begin{enumerate}

\item {\it Hamilton-Jacobi equations}

\begin{equation}\label{FE1}
g'=-2NK+{\mathcal{L}}_{X}g,
\end{equation}

\begin{equation}\label{FE2}
K'=-\nabla^{2}N+N(Ricc+kK-2K\circ K)+{\mathcal{L}}_{X}K-8\pi GN({\bf T}-\frac{tr_{\bf g}{\bf T}}{2}{\bf g}),
\end{equation}

\item {\it Constraint equations (energy and momentum respectively)}

\begin{equation}\label{FE3}
R-|K|^{2}+k^{2}=16\pi G\rho,
\end{equation}

\begin{equation}\label{FE4}
\nabla.K=-8 \pi G J,
\end{equation}

\item {\it Lapse equation} (deduced from equations above)

\begin{equation}\label{FE5}
-\Delta N+(4\pi G(\rho+3p)+|K|^{2})N=1.
\end{equation}

\end{enumerate}

\noindent The ${\bf T}$-term in the right hand side of equation \ref{FE2} must be thought to be restricted to $\Sigma$. Also
as usual $\rho={\bf T}(T,T)$, $J={\bf T}(T,.)$ and $p=\frac{({\bf T}_{ab})(g^{ab})}{3}$ is the average of the principal pressures. 
In equation \ref{FE4}, $\nabla.K=\nabla^{a}K_{ab}$ is the divergence and in equation \ref{FE2} 
it is $(K\circ K)_{ab}=K_{ac}K^{c}_{\ b}$. Finally the speed of light was taken to be $c=1$.

{\center \subsection{The Bel-Robinson energy and the space time curvature.}\label{subsec2.2}} 

\vspace{0.1cm}
We will measure the $L^{2}$ norm of the space time curvature relative to the CMC gauge. We will also need
to measure the $L^{2}$ norm of their time derivatives relative to the normal direction to the CMC foliation. 
There is a remarkable way to introduce them and it is by means of Weyl fields. Although we won't discuss Weyl
fields in detail as there are very accurate references on the subject (\cite{CK},\cite{A-M}), we will mention the most
used properties here and briefly elaborate on their conceptual importance as variables controlling the gravitational field.\\

\begin{Def} A Weyl field is a traceless $(4,0)$ space time 
tensor satisfying the symmetries of the curvature tensor ${\bf Rm}$. We will denote them by ${\bf W}_{abcd}$ or simply 
${\bf W}$.
\end{Def}

The Riemann tensor of a vacuum solution to the Einstein equations is a Weyl field that we will denote as 
${\bf Rm}={\bf W}_{0}$. Let $T$ be the normal field (future pointing) to the CMC foliation. Then 
$\bn_{T}^{i}{\bf W}_{0}={\bf W}_{i}$ are Weyl fields. Together with the {\it volume radius} (\cite{R}) and the $L^{2}$ norm
of the second fundamental form $K$ they are an important set of variables that control the gravitational field (i.e.
the metric ${\bf g}$ relative to the foliation) see \cite{R}. A central advantage for taking them as variables
is that they enjoy remarkable algebraic properties that simplifies the space time algebra considerably. We discuss the main formulae below. 
Given a Weyl tensor ${\bf W}$ define the left and
right duals $^{*}{\bf W}_{abcd}=\frac{1}{2}\epsilon_{ablm}{\bf W}^{lm}_{\ \ cd}$ and ${\bf W}^{*}_{abcd}={\bf W}_{ab}^{\ \ lm}
\frac{1}{2}\epsilon_{lmcd}$. Both are Weyl tensors, $^{*}{\bf W}={\bf W}^{*}$ and $^{*}(^{*}{\bf W})=-{\bf W}$. 
Define the current $J({\bf W})$ and its dual $J^{*}({\bf W})$ as

\begin{equation}
\bn^{a}{\bf W}_{abcd}=J_{abc}({\bf W}),
\end{equation}
\begin{equation}
\bn^{a}{\bf W}^{*}_{\ abcd}=J^{*}_{abc}({\bf W}).
\end{equation}

\noindent For the Riemann tensor in a vacuum solution to the Einstein equation we have $J=J^{*}=0$ due to the
Bianchi equations. This is a central fact that will be of fundamental importance latter. We also have

\begin{equation}
\bn_{[a}{\bf W}_{bc]de}=\frac{1}{3}\epsilon_{fabc}J^{*f}_{de}({\bf W}),
\end{equation}
\begin{equation}
\nabla_{[a}{\bf W}^{*}_{bc]}de=\frac{1}{3}\epsilon_{fabc}J^{f}_{de}({\bf W}).
\end{equation}

\noindent The $L^{2}$ norm with respect to the foliation will be defined through the Bel-Robinson tensor. Given a Weyl
field ${\bf W}$ its Bel-Robinson tensor is

\begin{equation}
Q_{abcd}({\bf W})={\bf W}_{alcm}{\bf W}_{b\ d}^{\ l\ m}+{\bf W}_{alcm}^{*}{\bf W^{*}}_{b\ d}^{\ l\ m}.
\end{equation}
    
\noindent It is symmetric and traceless in all pair of indices and for any pair of timelike vectors $T_{1}$ and $T_{2}$,
$Q(T_{1},T_{1},T_{2},T_{2})$ is positive if ${\bf W}\neq 0$ (\cite{CK}). In particular we define the $L^{2}$ 
norm of
${\bf W}$ with respect to the foliation as $Q(T,T,T,T)$. It is seen to be the $L^{2}$ norm of the electric and
magnetic fields of ${\bf W}$ defined through

\begin{equation}
E_{ab}({\bf W})={\bf W}_{acbd}T^{c}T^{d},
\end{equation}

\begin{equation}
B_{ab}({\bf W})=^{*}{\bf W}_{acbd}T^{c}T^{d}.
\end{equation}

\noindent i.e. $Q(T,T,T,T)=|E|^{2}+|B|^{2}$. They are symmetric, traceless and null on the $T$ direction. For 
the Riemann tensor in particular we have

\begin{equation}\label{de}
E_{ab}({\bf W}_{0})=Ricc_{ab}+kK_{ab}-K{a}^{\ c}K^{c}_{\ b},
\end{equation}

\noindent and

\begin{equation}
\epsilon_{ab}^{\ \ l}B_{lc}({\bf W}_{0})=\nabla_{a}K_{bc}-\nabla_{b}K_{ac}.
\end{equation}

\noindent The following formulae provide the components of a Weyl field with respect to the
CMC foliation in terms of the electric and magnetic fields ($i,j,k,l$ are spatial indices) 

\begin{equation}
{\bf W}_{ijkT}=-\epsilon_{ij}^{\ \ m}B_{mk}({\bf W}),\ ^{*}{\bf W}_{ijkT}=\epsilon_{ij}^{\ \ m}E_{mk}({\bf W}),
\end{equation}

\begin{equation}
{\bf W}_{ijkl}=\epsilon_{ijm}\epsilon_{kln}E^{mn}({\bf W}),\ ^{*}{\bf W}_{ijkl}=\epsilon_{ijm}\epsilon_{kln}B^{mn}({\bf W}).
\end{equation}

\noindent The divergence formula

\begin{equation}
\bn^{a}Q({\bf W})_{abcd}={\bf W}_{b\ d}^{\ m\ n}J({\bf W})_{mcn}+{\bf W}_{b\ c}^{\ m\ n}J({\bf W})_{mdn}+
\end{equation}

\begin{equation}
\ \ \ \ \ \ \ \ \ \ + ^{*}{\bf W}_{b\ d}^{m\ n}J^{*}({\bf W})_{mcn}+^{*}{\bf W}_{b\ c}^{m\ n}J^{*}(W)_{mcn},
\end{equation}

\noindent and therefore

\begin{equation}
\bn^{\alpha}Q({\bf W})_{\alpha TTT}=2E^{ij}({\bf W})J({\bf W})_{iTj}+2B^{ij}J^{*}({\bf W})_{iTj}
\end{equation}
\noindent gives the {\it Gauss equation}

\begin{equation}
\frac{\partial \int_{\Sigma}Q(T,T,T,T)dv_{g}}{\partial t}=-\int_{\Sigma}2NE^{ij}({\bf W})J({\bf W})_{iTj}+2NB^{ij}J^{*}({\bf W})_{iTj}
+3NQ_{abTT}\Pi^{ab}dv_{g},
\end{equation}

\noindent where $\Pi_{ab}=\bn_{a}T_{b}$ is the deformation tensor and plays a fundamental role in the tensor algebra. 
In terms of the electric and magnetic fields the components of $Q_{abTT}$ are written as

\begin{equation}
Q_{iTTT}=2(E\wedge B)_{i},
\end{equation}

\begin{equation}
Q_{ijTT}=-(E\times E)_{ij}-(B\times B)_{ij}+\frac{1}{3}(|E|^{2}+|B|^{2})g_{ij}.
\end{equation}

\noindent Controlling $J$ and $J^{*}$ in $L^{2}$ and $\Pi$ in $H^{2}$ is enough to control the
$L^{2}$ norm of the Weyl field. The following formulas are essential when it comes to get Sobolev estimates of
the Weyl field

\begin{equation}\label{eq1}
div E({\bf W})_{a}=(K\wedge B({\bf W}))_{a}+J_{TaT}({\bf W}),
\end{equation}
\begin{equation}\label{eq2}
div B({\bf W})_{a}=-(K\wedge E({\bf W}))+J^{*}_{TaT}({\bf W}),
\end{equation}
\begin{equation}\label{eq3}
curl B_{ab}({\bf W})=E(\bn_{T}{\bf W})_{ab}+\frac{3}{2}(E({\bf W})\times K)_{ab}-\frac{1}{2}kE_{ab}({\bf W})+J_{aTb}({\bf W}),
\end{equation}
\begin{equation}\label{eq4}
curl E_{ab}({\bf W})=B(\bn_{T}{\bf W})_{ab}+\frac{3}{2}(B({\bf W})\times K)_{ab}-\frac{1}{2}kB({\bf W})_{ab}+J^{*}_{aTb}({\bf W}),
\end{equation}

\noindent where the operations $\wedge,\ \times$ are defined as

\begin{equation}
(A\times B)_{ab}=\epsilon_{a}^{\ cd}\epsilon_{b}^{\ ef}A_{ce}B_{df}+\frac{1}{3}(A-B)g_{ab}-\frac{1}{3}(tr A)(tr B)g_{ab},
\end{equation}

\begin{equation}
(A\wedge B)_{a}=\epsilon_{a}^{\ bc}A_{b}^{\ d}B_{dc}.
\end{equation}

\noindent The equations \ref{eq1}-\ref{eq4} above are an example of the so called elliptic Hodge systems (\cite{CK}). In particular 
under basic regularity of the background metric they make possible to get elliptic estimates.

{\center \subsection{Scaling.}}

\vspace{0.1cm}
Scaling is the operation allowing us to speak like ``looking the system
at a particular scale". It is a different operation of than coordinate scaling, as scaling a solution
does changes the solution but scaling coordinate systems doesn't. Both transformations are however important when 
used simultaneously. \\

\begin{Def} Given a solution ${\bf g}$ to the Einstein equations, we  call $\lambda^{2}{\bf g}$ the
solution ${\bf g}$ at the scale of $\frac{1}{\lambda}$ and we call $\lambda$ the scale factor. 
\end{Def}

\noindent We say that a CMC state $(g,K)$ is cosmologically scaled (or normalized) if $k=-3$ or the same ${\mathcal{H}}=1$ as we 
will see the Hubble parameter ${\mathcal{H}}$ is equal to $\frac{-k}{3}$. 
Given a state $(g,K)$ that gives rise to a global solution ${\bf g}$ we can scale it as $\frac{k^{2}}{9}{\bf g}$ to transform
the original state $(g,K)(k)$ into a cosmologically normalized state $(\frac{k^{2}}{9}g,\frac{-k}{3}K)$. Therefore 
a state $(g,K)$ has a cosmological scale of $\frac{3}{-k}=\frac{1}{\mathcal{H}}$. Say $(g,K)(k)$ is a CMC state, and say $U$ is 
some space time tensor constructed out of ${\bf g}$ that we are looking at the $k$-slice. The corresponding
values of $U$ on the same slice when we cosmologically scale the state $(g,K)$ will be denoted with a tilde (either
above or next to it) say $\tilde{U}$ or $U^{\sim}$. Thus $\tilde{g}=\frac{k^{2}}{9}g$ and $\tilde{K}=\frac{-k}{3}K$. 
In a CMC flow $(g,K)(k)$ we can cosmologically scale the solution ${\bf g}$ at every $k$ getting thus a flow of normalized 
states $(\tilde{g},\tilde{K})(k)$. In the flat cone case the cosmologically scaled flow is just $(g_{H},-g_{H})(k)$ and
what we will call stability of the flat cone will be the stability of the cosmologically scaled solutions. In
general a space time tensor will scale as $\lambda^{s}U$ for some weight $s$, therefore $\tilde{U}$ will be
just $\tilde{U}=(\frac{-k}{3})^{s}U$. We will indistinctly use $\frac{-k}{3}$ or ${\mathcal{H}}$ as the scale factor $\lambda$.
The following table shows how some main tensors transform when ${\bf g}\rightarrow \lambda^{2}{\bf g}$.

\begin{center}
\begin{tabular}{|c|c|}
\hline
${\bf g}$&$\lambda^{2}{\bf g}$\\
$g$&$\lambda^{2}g$\\
$K$&$\lambda K$\\
$k$&$\frac{k}{\lambda}$\\
$N$&$\lambda^{2}N$\\
$\phi$&$\phi$\\
${\bf W}_{i}$&$\lambda^{-i+2}{\bf W}_{i}$\\
$Q_{i}$&$\lambda^{-(2i+1)}Q_{i}$\\
\hline
\end{tabular}
\end{center}   

\noindent $\phi$ is the Newtonian potential defined below.

{\center \section{Averaged evolution.}\label{sec3}}

{\center \subsection{Averaged cosmological parameters and the averaging map.}\label{subsec3.1}}

\vspace{0.1cm}
We define the geometric parameters, $a(k)$ (universe's radius), $\tau(k)$ (proper time), and ${\mathcal{H}}(k)$ (Hubble parameter)
in volume average. All those parameters reduce to the standard FL parameters when the solution is homogeneous and isotropic.\\

\begin{Def}\label{avepar} Given an arbitrary CMC solution we define the universe's radius at an instant of time $k$ as 
$a(k)=V_{g(k)}^{\frac{1}{3}}$. The volume-averaged proper time $\tau(k)$ is defined through

\begin{equation}
\frac{d\tau}{dk}=\frac{\int_{\Sigma}Ndv_{g}}{V}.
\end{equation}

\end{Def}

\noindent Recalling that in the FL models the Hubble parameter is defined as ${\mathcal{H}}=\frac{1}{a}\frac{d a}{d \tau}$ 
we compute

\begin{equation}
{\mathcal{H}}=\frac{1}{V^{\frac{1}{3}}}\frac{dV^{\frac{1}{3}}}{d\tau}=\frac{1}{V^{\frac{1}{3}}}\frac{dV^{\frac{1}{3}}}{dk}\frac{dk}{d\tau}=
\frac{1}{V^{\frac{1}{3}}}\frac{1}{3}V^{-\frac{2}{3}}(\int_{\Sigma}-Nkdv_{g})\frac{V}{\int_{\Sigma}Ndv_{g}}=\frac{-k}{3}.
\end{equation}

\noindent Thus in arbitrary solutions ${\mathcal{H}}=\frac{-k}{3}$. This expression is valid also locally in the following
sense: define the cube of the local radius as the volume element $dv_{g}(k)$, then the local Hubble parameter is
one third the logarithmic derivative of the volume element with respect to the proper time in the normal direction to
the CMC slice $k$. A direct computation gives for the local Hubble parameter ${\mathcal{H}}=\frac{1}{3dv_{g}}\frac{dv_{g}}{d\tau}=\frac{-k}{3}$. 

The Friedman-Lema\^itre equations take the form

\begin{enumerate}

\item \begin{equation}\label{FL1}
{\rm First\ FL\ equation:\ }{\mathcal{H}}^{2}=-\frac{\int_{\Sigma}Rdv_{g}}{6V}+\frac{\int_{\Sigma}(16\pi G\rho +|\hat{K}|^{2})dv_{g}}{6V},
\end{equation}

\item \begin{equation}\label{FL2}
{\rm Second\ FL\ equation:\ }\frac{a''}{a}=\frac{-\int_{\Sigma}{\mathcal{N}}(4\pi G(\rho+3p)+|\hat{K}|^{2})dv_{g}}{3V}.
\end{equation}

\end{enumerate}

\noindent Where ${\mathcal{N}}=\frac{N}{\bar{N}}$ (bar denotes volume-average) and has average equal to one. The derivatives, denoted with a prime
are proper time derivatives, i.e. $'=\frac{d}{d\tau}$. The first FL equation is just the volume average of the energy constraint

\begin{equation}
16\pi \rho=R-|\hat{K}|^{2}+\frac{2}{3}k^{2}.
\end{equation}

\noindent Observe that to make it look closer to the second FL equation, we can multiply the energy constraint before 
integrating by ${\mathcal{N}}$ and integrate thereafter to get

\begin{equation}
{\mathcal{H}}^{2}=\frac{\int_{\Sigma}{\mathcal{N}}Rdv_{g}}{6V}+\frac{\int_{\Sigma}{\mathcal{N}}(16\pi G\rho +|\hat{K}|^{2})dv_{g}}{6V}.
\end{equation}

To obtain the second FL equation we observe that

\begin{equation}\label{e1}
(\frac{a'}{a})'=\frac{a''}{a}-(\frac{a'}{a})^{2}=\frac{a''}{a}-{\mathcal{H}}^{2},
\end{equation}

\noindent and

\begin{equation}\label{e2}
(\frac{a'}{a})'=\frac{d{\mathcal{H}}}{d\tau}=-\frac{1}{3}\frac{dk}{d\tau}=-\frac{V}{3\int_{\Sigma}Ndv_{g}}.
\end{equation}

\noindent On the other hand integrating the Lapse equation \ref{FE5} we get 

\begin{equation}\label{e3}
\int_{\Sigma}N(4\pi G(\rho+3p)+|\hat{K}|^{2})dv_{g}=V-3{\mathcal{H}}^{2}\int_{\Sigma}Ndv_{g}.
\end{equation}

\noindent Equations \ref{e1}, \ref{e2}, \ref{e3} together give equation \ref{FL2}.

Let's restate the standard ${\mathcal{K}}=-1$ FL models on the light of the description given above for arbitrary
solutions. If the solution is ${\bf g}=-d\tau^{2}+a(\tau)^{2}g_{H}$ on a manifold $\field{R}\times \Sigma$ then $a(\tau)=(\frac{V}{V_{H}})^{\frac{1}{3}}$
where $V_{H}$ is the volume of $\Sigma$ with the hyperbolic metric $g_{H}$ and $V$ is the volume with the metric $a(\tau)^{2}g_{H}$. Our choice
of radius for arbitrary solutions has been instead $a(\tau)=V^{\frac{1}{3}}$, we will make this choice in equations \ref{1p} and \ref{2p} below. 
We recall too that in the standard FL models the energy density and pressures are a function only of $\tau$ and for that reason they coincide 
with their volume averages. Taking these facts into account the standard FL equations are 

\begin{enumerate}

\item \begin{equation}\label{1p}
{\mathcal{H}}^{2}= \frac{\int_{\Sigma}(16\pi G\rho)dv_{g}}{6V}-\frac{{\mathcal{K}}V_{H}^{\frac{2}{3}}}{a^{2}}.
\end{equation}

\item \begin{equation}\label{2p}
\frac{a''}{a}=\frac{-\int_{\Sigma}(4\pi G(\rho+3p)dv_{g}}{3V}.
\end{equation}

\end{enumerate}

\noindent Observe that in the FL equation \ref{FL1} instead of the curvature term 
$-{\mathcal{K}}V_{H}^{\frac{2}{3}}/a^{2}$ we have the term 

\begin{equation}
-\frac{\int_{\Sigma}Rdv_{g}}{6V}=-(\frac{\int_{\Sigma}Rdv_{g}}{6V^{\frac{1}{3}}})\frac{1}{a^{2}},
\end{equation}

\noindent where the first factor in the last term of the previous equations is scale invariant and therefore 
equal to $V_{H}^{\frac{2}{3}}$ for any metric 
scaled from the hyperbolic metric (so is close to it for any metric scaled from a metric close to a hyperbolic metric). 

In order to establish a mathematical definition of the averaging problem in cosmology we define the {\it averaging map} 
from arbitrary CMC solutions into Lorentzian manifolds in the following way.

\begin{Def} Given an arbitrary CMC solution ${\bf g}$ on $\field{R}\times \Sigma$ with $\Sigma$ a compact hyperbolic manifold
define the volume-averaged solution as the Lorentzian space ($\field{R}\times\Sigma$, ${\mathcal{A}}({\bf g})$) with 
${\mathcal{A}}({\bf g})=\bar{\bf g}=-d\tau^{2}+\frac{a(\tau)^{2}}{V_{H}^{\frac{2}{3}}}g_{H}$,
where $\tau$ and $a(\tau)$ are the averaged proper time and radius as given in definition \ref{avepar}. $g_{H}$ is the unique (up to diffeomorphism) 
hyperbolic metric that $\Sigma$ accepts.
\end{Def}  

\noindent It is essential in the definition above that, due to Mostow's rigidity, there is one hyperbolic metric up to diffeomorphism in a given
hyperbolic manifold. That makes the definition of ${\mathcal{A}}({\bf g})$ unambiguous. 

In rough terms the averaging problem for arbitrary solutions can be stated as to whether the averaged space 
${\mathcal{A}}({\bf g})$ is ``asymptotically in time close" to an exact ${\mathcal{K}}=-1$ FL solution with the ``averaged 
energy density and pressures" ``asymptotically in time close" to the energy density and pressures of the exact FL model. One
may also replace ``asymptotically in time close" simply by ``close" all along evolution. Physically that would be a more adequate
question to ask. This definition however faces various indefiniteness, we comment on them below. 

i) The first is to give a precise meaning to ``averaged energy and pressures" for
arbitrary solutions. We can
safely say what they are for the material fields, as material fields posses densities of pressures and energy, but it is not 
known what they are for the gravitational field, and presumably they can not be isolated as densities. 
The old question on how to define the gravitational energy which shows up all through General Relativity is present also 
here. A consensual definition of energy is the total ADM energy, a global term comprising the energetic content of a 
global system. Despite all the satisfactory the expression is, it is defined in asymptotically flat space-times and not 
in the context of cosmological solutions. We will argue in subsection \ref{subsec3.6} on the validity of the averaging problem 
, at least asymptotically in time, if it is assumed a compact and extended relative of the {\it weak cosmic censorship conjecture} of Penrose, 
conjecture stated for 
asymptotically flat space-times. Indeed we will analyze the averaging problem under the assumption that, under a particular model 
for matter at natural scales (the small structure), it happens that, generically, cosmological solutions evolve into a finite set 
(however large it may be) of asymptotically flat stationary solutions separating from each other, with gravitational radiation in 
between and if in addition we compute the ``averaged energy density" as the volume-average of the ADM energies of the stationary solutions 
plus the volume-average energy of the gravitational radiation in between. 
Both terms, as we shall see, can safely be computed. We will call the assumption above {\it assumption} (C). The extent as to whether 
this idealized assumption would be applicable to the actual universe in which we live at present times is 
not under consideration here. However I would like to point out
one aspect that immediately jumps out and that it would have to be addressed with care. Assuming that galaxies conform the 
individual stationary solutions, there is the issue to establish, due to the large dark halos extended over diameters
many times their visible diameters, where (if somewhere) and how far the individual galaxies (including their halos)
become asymptotically flat. This lack of asymptotic flatness on large neighborhoods around the visible galaxy is
manifested in the well known flat rotation curves of stars with large orbital radius. 

ii) A second problem in the rough definition of the averaging problem given above is to specify the 
equation of state of the exact FL solution from the original solution at natural scales. On the light of assumption (C)
there are two situations possible, a radiative regime, of universes filled only with gravitational radiation, and a 
massive regime, of universes where in addition to radiation there are massive compact object (the stationary solutions). 
Both regimes, that we will call {\it radiation} and {\it mass gap} respectively, deserve different technical analysis. 
We will discuss the radiation regime in rigor and detail below in section \ref{sec4}. The analysis of the mass gap regime 
is done in subsection \ref{subsec3.5}. Although rigorously deduced from the assumption (C), it lacks a precise determination on the decay of
the radiation term. We will return to this point later.

iii) A third problem is to define in a quantitative manner the notion of ``closeness" between the averaged space and
the exact FL solution. Precisely, we have to specify the scale in which the solutions are compared and a law for the asymptotic
relation between them.      

{\center \subsection{The Friedman-Lema\^itre equations and the Newtonian potential.}\label{subsec3.2}}

\vspace{0.1cm}
A remarkable fact about the averaging formalism is that the second FL equation can be written only in terms of the volume-average of the 
Newtonian potential $\bar{\phi}$ and consequently $a(\tau)$, ${\mathcal{H}}(\tau)$ and $z(\tau)$ are determined only from
$\bar{\phi}$.\\ 

\begin{Def} Define the Newtonian potential $\phi$ as $\phi=\frac{Nk^{2}}{3}-1$. It satisfies
the Poisson equation (Lapse equation)

\begin{equation}\label{P}
\Delta \phi = (4\pi G(\rho+3p)+|\hat{K}|^{2})+(4\pi G(\rho+3p)+|\hat{K}|^{2}+3H^{2})\phi,
\end{equation}

\end{Def}

\noindent or making $e=4\pi G(\rho+3p)+|\hat{K}|^{2}$ 

\begin{equation}
\Delta\phi = e+(e+3H^{2})\phi.
\end{equation}

\noindent From the Maximum principle it is seen that $-1\leq \phi \leq 0$. Observe too that $\phi$ is an absolute potential, i.e.
there is no ambiguity in the level of energy in its definition (as can be deduced from the unicity of solutions in equation \ref{P}) and 
observe also that it is scale invariant. As defined here the Newtonian potential
of course coincides with the usual Newtonian potential in the weak field Newtonian regime (when $p\approx 0$ and $K\approx 0$). 
Compare also equation \ref{P} with the usual Poisson equation in Newtonian dynamics

\begin{equation}
\Delta \phi=4\pi G\rho.
\end{equation}

\noindent Equation \ref{P} is fundamental to understand the dynamics of the gravitational field in general and its 
analysis extracts among other things the time at which Newtonian dynamics appears, i.e. when is that gravitation gets ruled by classical 
Newtonian potentials at large scales. 
A straightforward calculation gives

\begin{equation}
\frac{a''}{a}={\mathcal{H}}^{2}\frac{\bar{\phi}}{1+\bar{\phi}},
\end{equation}

\noindent or

\begin{equation}
\frac{{\mathcal{H}}'}{{\mathcal{H}}^{2}}=\frac{-1}{1+\bar{\phi}}.
\end{equation}

\noindent where $\bar{\phi}$ is the volume-average of $\phi$. This equation can be used to get an equation for ${\mathcal{H}}$ as a 
function of red shift $1+z=\frac{V^{\frac{1}{3}}}{V(z)^{\frac{1}{3}}}$ 
($V$ is the present volume and $V(z)$ is the volume at the corresponding red shift). The relation is

\begin{equation}
\frac{d\ln {\mathcal{H}}}{d\ln (1+z)}=\frac{1}{1+\bar{\phi}}.
\end{equation}

\noindent One also obtains

\begin{equation}
\frac{d\ln (1+z)}{d\tau}=-{\mathcal{H}}.
\end{equation}

\noindent Of course an estimation of $\bar{\phi}$ as a function of
$\tau$, $z$ or ${\mathcal{H}}$ is needed to make use of the equations above.

{\center \subsection{The CMC energy.}\label{subsec3.3}}

\vspace{0.1cm}
We would like to define a formal quantity on CMC states on a compact manifold $\Sigma$ analogous to the total ADM mass
of asymptotically flat space-times. Restate the first FL equation \ref{1p} in the form\\

\begin{equation}
1-(\frac{{\mathcal{V}}_{inf}}{{\mathcal{V}}})^{\frac{2}{3}}=\Omega_{m},
\end{equation}

\noindent where we have defined ${\mathcal{V}}_{inf}$ as the absolute infimum of the reduced 
volume ${\mathcal{V}}={\mathcal{H}}^{3}V(g,K)$ among
the set of all CMC states $(g,K)$. It is known \cite{F-M1},\cite{R} that if $\Sigma$ is hyperbolic ${\mathcal{V}}_{inf}=V_{H}$. 
$\Omega_{m}$ is defined as usual as $\Omega_{m}=\frac{8\pi G \rho}{3{\mathcal{H}}^{2}}$. Thus the
density of mass $\rho$ and the Hubble parameter ${\mathcal{H}}$ determine the deviation of the reduced volume from
its absolute infimum. If $\frac{8\pi G\rho}{3H^{2}}\sim 0$ we get in particular the approximation

\begin{equation}\label{energy}
{\mathcal{M}}\approx\frac{1}{4\pi G{\mathcal{H}}}({\mathcal{V}}-{\mathcal{V}}_{inf}).
\end{equation}

\noindent This remarkable equation, expresses the total mass ${\mathcal{M}}$ in terms only of ${\mathcal{H}}$, $G$, the total volume $V$ and the
topological invariant $V_{H}$. As we shall see in section \ref{subsec3.5} it holds too, asymptotically in time, for general models 
under assumption (C). Inspired on it and equation \ref{energy} we define the total CMC energy as

\begin{Def} Define the CMC global energy as 

\begin{equation}
E_{CMC}=\frac{1}{4\pi G{\mathcal{H}}}({\mathcal{V}}-{\mathcal{V}}_{inf}).\label{ECMC}
\end{equation}

\end{Def}

{\center \subsection{The ADM limit of the CMC energy: radiation.}\label{subsec3.4}}

\vspace{0.1cm}
Recall that the Hessian of the ADM energy around the flat Minkowski space-time state $g=g_{E}$ and $K=0$ ($g_{E}$ is the 
euclidean metric) is (see for instance \cite{J-B})\\

\begin{equation}\label{linen}
8\pi G\delta^{(2)}E_{ADM}=\frac{1}{4}\int_{\field{R}^{3}}|\nabla g'_{TT}|^{2}dv+\int_{\field{R}^{3}}|K'_{TT}|^{2}dv+
8\pi G\int_{\field{R}^{3}}\delta^{(2)}\rho dv,
\end{equation}

\noindent where $TT$ means transverse-traceless with respect to the flat metric $g_{E}$. The Hessian of the reduced 
volume ${\mathcal{V}}$ was calculated in
\cite{F-M1}. We include below a calculation of the Hessian of the CMC energy \ref{ECMC} based on their analysis for the 
sake of completeness and clarity. The Hessian of the CMC energy in the limit when $k\rightarrow 0$ is locally the same as equation \ref{linen}, 
the precise expression is

\begin{equation}\label{ADM}
8\pi G\delta^{(2)}E_{CMC}=\int_{\Sigma}|K'_{TT}|^{2}dv_{g}+\frac{1}{4}\int_{\Sigma}|\nabla g'_{TT}|^{2}dv_{g}-\frac{{\mathcal{H}}^{2}}{2}\int_{\Sigma}|g'_{TT}|^{2}dv_{g}+8\pi G\int_{\Sigma}\delta^{(2)}\rho dv_{g}.
\end{equation}

\noindent where the background state is $(\frac{9}{k^{2}}g_{H},\frac{3}{k}g_{H})$. We thus see the local 
vanishing of the third term on the right hand side when ${\mathcal{H}}\rightarrow 0$. Observe that the kinetic term 
$\frac{|\hat{K}|^{2}}{16\pi G}$ deduced from expression \ref{ADM} (there is an extra factor of a half when we read the energy from its Hessian)  
is consistent with the first and second FL equations in the radiation regime, where the densities of gravitational energy and pressure, 
are unequivocally identified with $\rho_{G}=p_{G}=\frac{|\hat{K}|^{2}}{16\pi G}$. Note however that the first term in \ref{ADM} doesn't form part of the effective densities of gravitational energy and 
pressure in equation \ref{e2} and therefore doesn't influence the universe's deceleration, instead it is part of the
curvature term in the first FL equation.       
 
The calculation of the Hessian is as follows. In terms of conformal variables, a state $(g,K)$ is written as

\begin{equation}
g_{ab}=\varphi^{4}g_{Y,ab},
\end{equation}

\begin{equation}
K^{ab}=\varphi^{-10}\hat{K}_{Y}^{ab}+\frac{k}{3}\varphi^{-4}g_{Y}^{ab}.
\end{equation}

\noindent Where $g_{Y}$ is a Yamabe metric of constant scalar curvature $R_{Y}=-6\frac{k^{2}}{9}$ and $\hat{K}_{Y}$ is a transverse traceless
tensor with respect to $g_{Y}$. The conformal factor $\varphi$ must satisfy the Lichnerowicz equation

\begin{equation}
\Delta \varphi +\frac{k^{2}}{12}(\varphi-\varphi^{5})+\frac{|\hat{K}_{Y}|_{Y}^{2}}{8}\varphi^{-7}+2\pi G\rho \varphi^{5}=0.
\end{equation}

\noindent We will take derivatives along a path $(g,K)(\lambda)$ with $(g,K)(0)=(\frac{9}{k^{2}}g_{H},\frac{3}{k}g_{H})$, which in turn 
can be seen as a path $(g_{Y},K_{Y},\varphi)(\lambda)$. Note that $\varphi(0)=1$. Recalling the derivative of the Laplacian (\cite{Besse})

\begin{equation}
-(\Delta')f=<\nabla^{2}f,g'>-<\nabla f,\delta h>-\frac{1}{2}<\nabla f,d tr_{g} g'>,
\end{equation}

\noindent the first derivative at $\lambda=0$ of the Lichnerowicz equation is (we are assuming $\delta^{(1)}\rho=0$)

\begin{equation}
\Delta \varphi'-\frac{k^{2}}{3}\varphi'=0.
\end{equation}

\noindent which shows that $\varphi'(0)=0$ identically. Using that fact we get

\begin{equation}\label{11}
V''(0)=(\int_{\Sigma}\varphi^{6}dv_{g})''=6\int_{\Sigma} \varphi''dv_{g(0)}+\int_{\Sigma}dv_{g_{Y}}''.
\end{equation}

\noindent Integrating the Lichnerowicz equation and differentiating the integral equation twice gives 

\begin{equation}\label{3}
\frac{8k^{2}}{3}\int_{\Sigma}\varphi''dv_{g(0)}=2\int_{\Sigma}|\hat{K}_{Y}'|^{2}dv_{g(0)}+16\pi G\int_{\Sigma}\delta^{(2)}\rho dv_{g(0)},
\end{equation}

\noindent from which we get

\begin{equation}
6\int_{\Sigma}\varphi''dv_{g(0)}=\frac{9}{2k^{2}}\int_{\Sigma}|\hat{K}'|^{2}dv_{g(0)}+\frac{9}{2k^{2}}8\pi G\int_{\Sigma}\delta^{(2)}\rho dv_{g(0)}.
\end{equation}

\noindent Now let's compute the second term in equation \ref{11}. First we note that 

\begin{equation}\label{2}
dv_{g_{Y}}''=(\frac{tr_{g_{Y}}g_{Y}''}{2}-\frac{|g'_{Y}|^{2}}{2} + (\frac{tr_{g}g'}{2})^{2})dv_{g_{Y}}.
\end{equation}

\noindent To compute $tr_{g_{Y}}g''_{Y}$ we will use the variation formula for the scalar curvature. As the metrics $g_{Y}$ are
Yamabe of scalar curvature $-6\frac{k^{2}}{9}$ the derivative in $\lambda$ of $R_{Y}$ is zero pointwise, precisely (\cite{Besse})

\begin{equation}\label{vR}
R'=-\Delta (tr_{g_{Y}}g'_{Y})+\delta\delta g'_{Y}-<Ric,g'>=0.
\end{equation}

\noindent Integrating we get

\begin{equation}
\int_{\Sigma}<Ric,g'_{Y}>dv_{g_{Y}}=0,
\end{equation}

\noindent for all $\lambda$. Differentiating again at $\lambda=0$ we get

\begin{equation}\label{122}
\int_{\Sigma}(<Ric',g'_{Y}> + <Ric,g''_{Y}> + (Ric_{ab})(g'_{Y,cd})(g^{ac}_{Y})'(g^{bd}_{Y}) + (Ric_{ab})(g'_{Y,cd})(g^{ac}_{Y})(g^{bd}_{Y})') dv_{g(0)}
\end{equation}
 
\noindent The Ricci curvature at $\lambda=0$ is $Ric=\frac{-2k^{2}}{9}g_{H}$. Also the functional derivative 
of Ricci is

\begin{equation}
Ric'=\frac{1}{2}\Delta_{L}g'-\delta^{*}(\delta g')-\frac{1}{2}\nabla\nabla(tr_{g}g').
\end{equation}

\noindent Observe that from equation \ref{vR} we have $tr_{g(0)}g'(0)=0$. $\Delta_{L}$ is the Lichnerowicz laplacian and has the 
expression \cite{Besse}

\begin{equation}
\Delta_{L}T_{ab}=\nabla^{*}\nabla T_{ab}+(Ric_{ac}T^{c}_{\ b}+Ric_{bc}T^{c}_{\ a})-(Rm_{acbd}T^{cd}+Rm_{bcad}T^{dc}).
\end{equation}

\noindent Using both facts and also that $g'$ is taken to be transverse we get from equation 
\ref{122} that

\begin{equation}\label{12}
\int_{\Sigma}tr_{g_{Y}}g''_{Y}dv_{g(0)}=2\int_{\Sigma}|g'_{Y}|^{2}dv_{g(0)}+\frac{9}{4k^{2}}\int_{\Sigma}<g'_{y},\Delta_{L}g'_{Y}>dv_{g(0)}.
\end{equation}

\noindent To compute the Lichnerowicz laplacian we remember that the sectional curvature of $g(0)$ is 
$-\frac{k^{2}}{9}$, therefore

\begin{equation}
\Delta_{L}g'_{Y}=\nabla^{*}\nabla g'_{Y}-\frac{6k^{2}}{9}g'_{Y}.
\end{equation}

\noindent at $\lambda=0$. Using the previous equation in equation \ref{12}, we get the result of equation \ref{ADM} 
after putting together equations \ref{2}, \ref{3}, \ref{11}.

{\center \subsection{The long time ADM limit of the CMC energy: radiation and mass gap.}\label{subsec3.5}}

\vspace{0.1cm}
In this subsection we will introduce assumption (C) and show how, under that assumption, the CMC energy converges assymptotically in 
time to the sum of the ADM masses of the emerging stationary solutions plus a radiative term of the radiation in between. The analysis 
will lead us to argue in subsection \ref{subsec3.6} on the validity of the averaging problem in cosmology under assumption (C) and
asymptotically in time.
First we recall the definition of asymptotically flat stationary solution.\\

\begin{Def} (\cite{H}, pg 16) A maximal ($k=0$) initial data set $(g,K,N,X)$ is a stationary asymptotically flat data iff 

\begin{enumerate}

\item 

\begin{enumerate}

\item $g_{00}=-(1-\frac{2M}{r})+O(r^{-2})$,
\item $g_{ij}=(1+O(r^{-1}))\delta_{ij}+O(r^{-2})$,
\item $g_{0i}=-\epsilon_{ijk}\frac{4S^{j}}{r^{3}}x^{k}+O(r^{-3})$.

\end{enumerate}

\item it satisfies the stationary vacuum Einstein equations $\dot{g}=\dot{K}=0$.

\end{enumerate}

\end{Def}

\begin{figure}
\centering
\includegraphics[width=100mm,height=50mm]{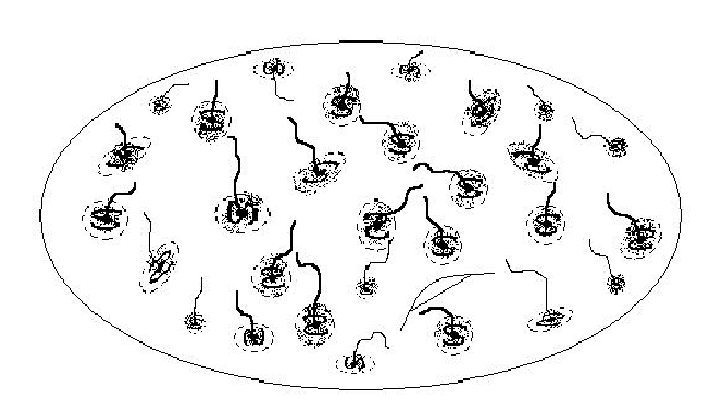}
\caption{A schematic representation of the cosmological scale of a universe satisfying assumption (C) after 
sufficiently long time. The
emerging stationary solutions are represented with a galactic symbol enclosed in a dashed circle representing 
the outer spheres. The tails coming out from the inside of the emerging stationary solutions represent the large tubes
developing inside possible black holes.}\label{fig1} 
\end{figure}

Now we state the definition of assumption (C). A schematic representation of a space-time (at a given time) 
satisfying assumption (C) is given in figure \ref{fig1}.

\begin{Def} A long time CMC solution satisfies the assumption (C) iff:

\begin{enumerate}

\item (emergence of isolated stationary solutions) after a sufficiently large time there is a finite set of pairs of 
two-spheres (inner and outer) with constant mean curvatures $2/L_{0}$ and $2/L(t)$ respectively, varying 
continuously in time ($t=1/{\mathcal{H}}$) such that, inside the annulus in between, the unscaled flow $(g,K,N,X)$ decays 
in the $C^{1}$ norm into a stationary solution $(g_{0},K_{0},N_{0},X_{0})$. At the outer spheres, $|\nabla \phi-\nabla \phi_{0}|\leq
\frac{C}{L(t)^{2}t^{1+\epsilon}}$.  

\item (the inside of the inner spheres) after a sufficiently long time the volume of the inside of the inner spheres 
grows no faster than $t^{1-\epsilon}$. 

\item (emergence of the radiative region) after a sufficiently long time the cosmologically normalized 
flow $(\tilde{g},\tilde{K},\tilde{N},\tilde{X})$, decays uniformly in $C^{1}$, over the exterior region to the outer spheres into the flat cone state $(g_{H},-g_{H},1/3,0)$. 

\item (boundedness of the CMC energy) $\frac{dE}{d t}\leq \frac{C}{t^{2+\epsilon}}\rightarrow 0$.

\end{enumerate}

\end{Def}

\noindent Some remarks are in order. The interior radius $L_{0}$ is fixed. The exterior radius $L(t)$ grows monotonically 
but less than $t$: $\lim_{t\rightarrow \infty}\frac{L(t)}{t}=0$, in such a way that at cosmological scales the outer spheres
get smaller and smaller in size. Similarly, the rate at which the solution over the annulus decays into the stationary 
solution and the rate at which the solution on the exterior region decays into the flat cone solution are left unspecified here. 
Item four is a global condition that
complements the absence of explicit decaying rates in assumption (C). In the CMC flow, the interior regions of black holes are 
expected to evolve 
as tubes of increasingly large size, and therefore increasing volume. Item two gives a bound on its grow in the case they form. 
We want to stress that all these conditions are tentative and are not intended to be conjectural. Neither we  conjecture a sort of
assumption (C) to hold generically. The introduction of assumption (C), we believe, provides an starting point in the study of the 
averaging problem directly from the small structure of exact solutions. All these problems are, however, difficult problems in the field. 
Section \ref{sec4} is an attempt to clarify these issues in pure radiative solutions.

Now let's see how the CMC energy behaves under assumption (C). The second FL equation in terms of the CMC energy is

\begin{equation}\label{EC}
\frac{d E_{CMC}}{d \sigma}=-\int_{\Sigma}((\rho+3p)+\frac{|\hat{K}|^{2}}{4\pi G})(1+\phi)dv_{g}+E_{CMC}=3{\mathcal{H}}^{2}\int_{\Sigma}\phi dv_{g}+E_{CMC},
\end{equation}

\noindent where $\sigma=-\ln -k$ is the logarithmic time. From item four and equation \ref{EC} the CMC energy converges 
to the term $-3{\mathcal{H}}^{2}\int_{\Sigma}\phi dv_{g}$ with a difference bounded by $C/t^{1+\epsilon}$. Now let's separate the 
region of integration into the inside of the outer spheres and its outside. Using the Poisson equation \ref{P} we get 

\begin{equation}\label{final}
E_{CMC}=\int_{S_{out}}<\nabla\phi,n_{out}>dA+3{\mathcal{H}}^{2}\int_{\Omega_{int}}\phi dv_{g}+\int_{\Omega_{ext}}
((\rho+3p)+\frac{|\hat{K}|^{2}}{4\pi G})(1+\phi)dv_{g}+O(t^{-(1+\epsilon)}),
\end{equation}

\noindent where $\Omega_{int}$ is the interior of the outer spheres and $\Omega_{ext}$ its exterior. Due to item two
in assumption (C), the second term on the right hand side of equation \ref{final} is an $O(t^{-(1+\epsilon)})$. The boundary 
term approach with an error $O(t^{-(1+\epsilon)})$ to the sum of the ADM masses of the emerging stationary
solutions. We can identify the third term on the right hand side of equation \ref{final} is the radiative term because by item three 
$\phi\rightarrow 0$ pointwise on $\Omega_{ext}$ and the radiation terms from matter and gravitation decouple. Thus we get
 
\begin{equation}
E_{CMC}\approx {\mathcal{M}}+{\mathcal{R}},
\end{equation}

\noindent the total ADM mass plus the radiation energy. This is the same equation as \ref{energy} with the additional 
radiative term. A remark has to be said about the radiative term. In an asymptotically flat context the ADM energy is a conserved
quantity, therefore the radiative contribution to energy measured as the difference between the asymptotically Bondi energy and 
the ADM energy would be a definite non-zero amount. In other words there is a definite amount of radiative energy that forms part 
of the ADM energy. In our context, that amount would form part of the radiative term ${\mathcal{R}}$. Further work is needed to
show that, indeed there may exist a non vanishing residual radiative energy in the ${\mathcal{R}}$ term.     

We will use the total CMC energy in section \ref{sec4} to give a rigorous estimation of the gravitational 
energy in the long time for radiative solutions.   

{\center \subsection{The averaging problem in cosmology.}\label{subsec3.6}}

\vspace{0.1cm}
We will discuss here the implications of assumption (C) for the averaging problem in cosmology. Noting that 
${\mathcal{N}}=\frac{1+\phi}{1+\bar{\phi}}$ we rewrite the second FL equation in the form\\ 

\begin{equation}\label{FL22}
\frac{a''}{a}=\frac{-\int_{\Sigma}(4\pi G(\rho+3p)+|\hat{K}|^{2})(1+\phi)dv_{g}}{3(1+\bar{\phi})V}={\mathcal{H}}^{2}\frac{\bar{\phi}}{1+\bar{\phi}}.
\end{equation}

\noindent with

\begin{equation}
\bar{\phi}=\frac{-\int_{\Sigma}(4\pi G(\rho+3p)+|\hat{K}|^{2})(1+\phi)dv_{g}}{3{\mathcal{H}}^{2}V}
\end{equation}

The integrand is the same as in equation \ref{EC} therefore we can decompose the integration as we did in
equation \ref{final}. Note that if in equation \ref{EC} we write 
$3{\mathcal{H}}^{2}\int_{\Sigma}\phi dv_{g}=-3t{\mathcal{V}}\bar{\phi}$, we get because of item four in assumption (C)
and the fact that the reduced volume is monotonically decreasing and bounded below by $V_{H}$ that 
$\bar{\phi}=\frac{E_{CMC}}{-3t{\mathcal{V}}}+O(t^{-(2+\epsilon)})=O(1/t)$. 
This gives the estimation that the factor $1+\bar{\phi}$ in the denumerator of equation \ref{FL22} behaves as $1+O(t^{-1})$.  
All together gives

\begin{equation}\label{AVE}
\frac{1}{a}\frac{d^{2} a}{d\tau^{2}}=-\frac{4\pi G(\bar{\mathcal{M}}_{ADM}+
\bar{\mathcal{R}})}{3-4\pi G{\mathcal{H}}^{-2}(\bar{\mathcal{M}}_{ADM}+\bar{{\mathcal{R}}})}+O(t^{-(4+\epsilon)})=
-\frac{4\pi G(\bar{\mathcal{M}}_{ADM}+\bar{\mathcal{R}})}{3}+O(t^{-(3+\epsilon)})
\end{equation} 

\noindent where $\bar{\mathcal{M}}_{ADM}$ is the volume average of the sum of the ADM masses of the emerging stationary solutions, and
$\bar{\mathcal{R}}=\bar{\rho}_{rad}+3\bar{p}_{rad}+\bar{\rho}_{G}+3\bar{p}_{G}$ where $\bar{\rho}_{rad}$, 
$\bar{p}_{rad}$ and $\bar{\rho}_{G}$, $\bar{p}_{G}$ are the volume average of the energy and pressure densities of 
material and gravitational radiation respectively. Equation \ref{AVE} is a differential equation in $\tau$, however
the estimate on its right hand side is in terms of $t=1/{\mathcal{H}}$. We thus complement this equation with a differential equation
for $\tau$ as a function of $t$. From the defining equation of $\tau$ we get the equation  

\begin{equation}\label{taut}
\frac{d\tau}{d t}=1+\bar{\phi}=1-\frac{4\pi G}{3}t^{2}(\bar{\mathcal{M}}_{ADM}+\bar{\mathcal{R}}).
\end{equation}

\noindent Equations \ref{AVE} and \ref{taut} are the main equations for the averaging problem under assumption (C) and
asymptotically in time. We remark that still the Einstein equations have to be used in full, to provide an estimation of the
radiative term $\bar{\mathcal{R}}$. The next section intends to provide these estimates in the case ${\mathcal{M}}_{ADM}=0$, i.e.
a purely radiative solution. 

{\center \section{Long time smoothing and estimates on the gravitational energy: radiation.}\label{sec4}}

We will use the notation $H^{s}$ for the Sobolev space with $s$ derivatives and $H^{s}_{g_{H}}$ for the Sobolev
space where the norms and covariant derivatives are calculated via $g_{H}$ (see \cite{R}). We will prove here Theorem 
\ref{Smooth}. The proof is a natural extension of the analysis in \cite{A-M}.\\ 

\begin{T} {\rm (Expansive smoothing and energy estimates).} Let $\Sigma$ be a compact and rigid hyperbolic manifold. 
There is an $\epsilon>0$ such that the Einstein  CMC flow of a cosmologically scaled initial state (i.e. with ${\mathcal{H}}=1$) $(g,K)$ 
with ${\mathcal{V}}-{\mathcal{V}}_{inf}\leq \epsilon$
and $\tilde{\mathcal{E}}_{1}\leq \epsilon$ has the following long time properties (take $t=\frac{1}{\mathcal{H}}$):

\begin{enumerate}

\item The limit $lim_{t\rightarrow \infty} t^{3}Q_{0}$ is finite and greater than zero.

\item there are $n_{i}\geq 0$ such that $\lim_{t\rightarrow \infty} \frac{t^{2i+3}}{(\ln t)^{n_{i}}}Q_{i}\leq\infty$ for $i\geq 1$.

\item for given $\gamma>0$, $\int_{t}^{\infty}\frac{\int_{\Sigma}|\hat{K}|^{2}dv_{g}}{u}du\geq Ct^{-(2+\gamma)}$.

\item $|\hat{K}|^{2}\leq Ct^{-4}$ pointwise (not volume averaged).

\end{enumerate}

\noindent In particular the cosmologically scaled flow of a $H^{i}\times H^{i-1}$ state (for any $i\geq 1$) as in the
hypothesis above 
converges in $H^{i}\times H^{i-1}$ to the canonical flat cone state $(g,K)=(g_{H},-g_{H})$.

\end{T} 

\noindent{\bf Proof of theorem \ref{Smooth}}. We start by recalling a result from \cite{A-M} that will be useful
to prove items $1$ and $2$ in theorem \ref{Smooth}. 

\begin{Lem} {\rm \ref{Smooth}}\label{Estimate} Let $\Sigma$ be a compact and rigid hyperbolic manifold. There are  
$C$ and $\epsilon_{0}$ such that if a cosmologically normalized CMC state $(g,K)$, where $g$ is harmonic with respect to $g_{H}$, is $\epsilon$-close 
to $(g_{H},-g_{H})$ in the $H^{3}_{g_{H}}\times H^{2}_{g_{H}}$ topology, with $\epsilon\leq \epsilon_{0}$ then 
there is a constant $C$ (dependent on $\epsilon_{0}$) such that

\begin{equation}
C^{-1}\tilde{\mathcal{E}}_{1}\leq (\|g-g_{H}\|^{2}_{H^{3}_{g_{H}}}+\|K+g_{H}\|^{2}_{H^{2}_{g_{H}}})\leq C\tilde{\mathcal{E}}_{1}.
\end{equation}

\noindent We get therefore the elliptic estimate for the Newtonian potential $\phi=\hat{N}=\frac{k^{2}N}{3}-1$ from the 
lapse equation 

\begin{equation}
\|\hat{N}\|_{H^{2}_{g_{H}}}\leq C\|\hat{K}\|_{H^{2}_{g_{H}}}\|\hat{K}\|_{L^{2}_{g_{H}}}\leq C\tilde{\mathcal{E}}_{1},
\end{equation}

\noindent and
\begin{equation}
\|\hat{N}\|_{H^{3}_{g_{H}}}\leq C\|\hat{K}\|_{H^{2}_{g_{H}}}\|\hat{K}\|_{H^{1}_{g_{H}}}\leq C\tilde{\mathcal{E}}_{1}.
\end{equation}

\end{Lem}

To extract conclusions on the decay of the Sobolev norms of the cosmologically normalized states we will make use
of the fact proved in \cite{R} that  under the conditions of the last lemma, $\epsilon_{0}$ and 
$\tilde{\mathcal{E}}_{i-1}$ controls the difference of the states in $H^{i}_{g_{H}}\times H^{i-1}_{g_{H}}$ with respect 
to the background state $(g_{H},-g_{H})$ states at zero, i.e. the derivatives tend to zero in $L^{2}_{g_{H}}$ 
as $\epsilon_{0}$ and $\tilde{\mathcal{E}}_{i}$ tend to zero.

\noindent {\it Item 1}. The Gauss equation gives the following inequality for the evolution of the first order
cosmologically normalized Bel-Robinson energy (\cite{A-M})

\begin{equation}
\frac{d\tilde{\mathcal{E}}_{1}}{d\sigma}\leq -2\tilde{\mathcal{E}}_{1}+C\tilde{\mathcal{E}}_{1}^{\frac{3}{2}}.
\end{equation}

\noindent with $c$ a constant greater than zero. It follows therefore that $\tilde{\mathcal{E}}_{1}$ decays faster than the solution $x(\sigma)$ to the
following ordinary differential equation and same initial condition

\begin{equation}
x'=-2x+cx^{\frac{3}{2}}.
\end{equation}

\noindent This is a Bernoulli type of equation that can be solved by making the change of variables $v=x^{-\frac{1}{2}}$
which gives the differential equation

\begin{equation} 
v'=v-\frac{c}{2},
\end{equation}

\noindent having the solution $v=\frac{1}{2}+Ae^{\sigma}$. This implies that 

\begin{equation}
x=\frac{x(\sigma_{0})e^{-2(\sigma-\sigma_{0})}}{(\frac{c}{2}(e^{-(\sigma-\sigma_{0})}-1)x(\sigma_{0})^{\frac{1}{2}}+1)^{2}},
\end{equation}

\noindent which results in the following decay of $\tilde{\mathcal{E}}_{1}$

\begin{equation}
\tilde{\mathcal{E}}_{1}\leq \frac{\tilde{\mathcal{E}}_{1}(\sigma_{0})e^{-2(\sigma-\sigma_{0})}}
{(\frac{c}{2}(e^{-(\sigma-\sigma_{0})}-1)\tilde{\mathcal{E}}_{1}(\sigma_{0})^{\frac{1}{2}}+1)^{2}}.
\end{equation}

\noindent Observe that if $\sigma_{0}$ is big enough then we get the bound

\begin{equation}
\tilde{\mathcal{E}}_{1}\leq \frac{\tilde{\mathcal{E}}_{1}(\sigma_{0})e^{-2(\sigma-\sigma_{0})}}{4}.
\end{equation}

\noindent Now we prove item $1$ in
theorem \ref{Smooth}. From the Gauss equation and Lemma \ref{Estimate} and the above estimate for 
$\tilde{\mathcal{E}}_{1}$ we get an evolution equation for $\tilde{Q}_{0}$ of the form

\begin{equation}\label{form}
\frac{d\tilde{Q}_{0}}{d\sigma}=-2\tilde{Q}_{0}+h(\sigma),
\end{equation}

\noindent where $h(\sigma)$ is a function which is bounded in absolute value by

\begin{equation}
|h(\sigma)|\leq C\tilde{\mathcal{E}}_{1}^{\frac{3}{2}}(\sigma_{0})e^{-3(\sigma-\sigma_{0})}.
\end{equation}

\noindent Therefore we get the following expression for $\tilde{Q}_{0}$

\begin{equation}
\tilde{Q}_{0}=e^{-2(\sigma-\sigma_{0})}(\tilde{Q}_{0}(\sigma_{0})+e^{-2\sigma_{0}}\int_{\sigma_{0}}^{\sigma}h(u)e^{2u}du),
\end{equation}

\noindent Clearly the integral in $h$ has a limit when $\sigma\rightarrow \infty$. If the term in parenthesis 
on the right hand side has limit different than zero then we are done as then

\begin{equation}
\lim_{\sigma\rightarrow \infty}\frac{\tilde{Q}_{0}}{e^{-2\sigma}}>0.
\end{equation}

\noindent Let's see that the limit cannot be zero. If that happens then we have for all $\sigma$

\begin{equation}
\tilde{Q}_{0}(\sigma)=-e^{-2\sigma}\int_{\sigma}^{\infty}h(u)e^{2u}du.
\end{equation}

\noindent The integral is negative for all $\sigma$ ($\tilde{Q}_{0}$ is positive) and goes to zero as 
$\sigma\rightarrow \infty$. Then there is a diverging sequence $\{\sigma_{i}\}$ such that for all 
$\sigma\geq \sigma_{i}$ we have

\begin{equation}
-\int_{\sigma}^{\infty}h(u)e^{2u}du\leq -\int_{\sigma_{i}}^{\infty}h(u)e^{2u}du
\end{equation}

\noindent making then

\begin{equation}
\tilde{Q}_{0}(\sigma)\leq \tilde{Q}_{0}(\sigma_{i})e^{-2(\sigma-\sigma_{i})},
\end{equation}

\noindent for all $\sigma\geq \sigma_{i}$. Using again the Gauss equation, Lemma \ref{Estimate} and the estimate
above we get an evolution equation for $\tilde{Q}_{0}(\sigma)$ of the same form as in equation \ref{form} with
$h$ instead bounded in absolute value by 
$C\tilde{\mathcal{E}}_{1}(\sigma_{i})^{\frac{1}{2}}\tilde{Q}_{0}(\sigma_{i})e^{-3(\sigma-\sigma_{i})}$. It thus
gives an expression for $\tilde{Q}_{0}$ of the form

\begin{equation}\label{f}
\tilde{Q}_{0}(\sigma)=\tilde{Q}_{0}(\sigma_{i})e^{-2(\sigma-\sigma_{i})}(1+
e^{-2\sigma_{i}}\int_{\sigma_{i}}^{\sigma}\frac{h(u)e^{2u}}{\tilde{Q}_{0}(\sigma_{i})}du).
\end{equation}

\noindent To see that $\lim_{\sigma\rightarrow \infty}\tilde{Q}_{0}e^{2\sigma}>0$ we note the following bound
for the integral term in the equation \ref{f} above

\begin{equation}
|e^{-2\sigma_{i}}\int_{\sigma_{i}}^{\infty}\frac{h(u)e^{2u}}{\tilde{Q}_{0}(\sigma_{i})}du|\leq C\tilde{\mathcal{E}}_{1}(\sigma_{i})^{\frac{1}{2}},
\end{equation}

\noindent which tends to zero as $\sigma_{i}\rightarrow \infty$. This is a contradiction, thus the limit must be 
positive.  

\noindent {\it Item 2}. Now we prove item $2$. By induction we will be able to get an equation for $\tilde{\mathcal{E}}_{i}(\sigma)$
of the form

\begin{equation}\label{formula}
\tilde{Q}_{i}'=-(2+h'(\sigma))\tilde{Q}_{i}+h(\sigma)\tilde{Q}_{i}^{\frac{1}{2}},
\end{equation}

\noindent where $h'(\sigma)$ and $h(\sigma)$ are functions bounded in absolute value by $C'\sigma^{n'}e^{-\sigma}$ and $C\sigma^{n}e^{-\sigma}$ 
for some $C',C$ and $n',n$ constants. It follows after making the change of variable $v=\tilde{Q}_{i}^{\frac{1}{2}}$ that 
$\tilde{Q}_{i}$ can be bounded by an expression of the form

\begin{equation}\label{energ}
\tilde{Q}_{i}\leq C\sigma^{2(n+1)}e^{-2\sigma},
\end{equation}

\noindent for some constant $C$.

\begin{Lem}\label{induction} Suppose that a solution to the CMC flow $(g,K)$ has 

\begin{equation}\label{asymptotics}
\tilde{Q}_{j}(\sigma)\leq C_{j}\sigma^{n_{j}}e^{-2\sigma},
\end{equation}

\noindent for $j=0,\ldots,i\geq 1$, then $\tilde{Q}_{i+1}$ satisfies an equation of the form 
\ref{formula} and therefore satisfies an asymptotic of the form \ref{asymptotics} for $j=i+1$.
\end{Lem}  

\noindent{\bf Proof:} We start with the differential inequality for $\tilde{Q}_{i}$. Make $\beta=\frac{-3}{k}$.
Then $Q_{i}(k)=\lambda^{(2i-1)}Q_{i}(k)$, and therefore

\begin{equation}
\frac{d\tilde{Q}_{i}}{d\sigma}=\frac{3}{\beta}\frac{d\tilde{Q}_{i}}{dk}=
\frac{3}{\beta}((2i+1)\frac{\beta^{2i+2}}{3}Q_{i}+\beta^{2i+1}\frac{dQ_{i}}{dk}).
\end{equation}

\noindent A useful trick for the calculations that follow is to write 

\begin{equation}
\beta^{2i+1}\frac{dQ_{i}}{dk}=\beta\frac{dQ_{i}(\beta^{-2}{\bf g})}{d(\beta k)}
\end{equation}

\noindent where the $\beta$ inside the derivative on the right hand side is taken constant equal to its
value at the time of differentiation. Thus we are calculating the $k$-derivative of the cosmologically scaled
solution at $k=-3$. Putting all together we get

\begin{equation}\label{31}
\frac{d\tilde{Q}_{i}}{d\sigma}=(2i+1)\tilde{Q}_{i}+3\frac{dQ_{i}(\beta^{-2}{\bf g})}{d(\beta k)}.
\end{equation}

\noindent We are going to study the derivatives $\frac{d Q_{i}}{dk}$ of perturbation of the canonical flat
cone state $(g_{H},-g_{H})$ at $k=-3$. From the Gauss equation we have

\begin{equation}\label{Gaa}
\frac{dQ_{(i)}}{dk}=-3\int_{\sigma}NQ_{(i)abTT}{\Pi}^{ab}dv_{g}-\int_{\Sigma}2N(E_{(i)}^{ab}J_{(i)aTb}+B_{(i)}^{ab}J^{*}_{(i)aTb})dv_{g},
\end{equation} 

\noindent therefore 

\begin{equation}
3\beta^{2i+1}\frac{dQ_{i}}{dk}=-9\int_{\sigma}\tilde{N}\tilde{Q}_{(i)ab\tilde{T}\tilde{T}}{\tilde{\Pi}}^{ab}dv_{\tilde{g}}-
\int_{\Sigma}6\tilde{N}(\tilde{E}_{i}^{ab}\tilde{J}_{(i)a\tilde{T}b}+\tilde{B}_{i}^{ab}\tilde{J}^{*}_{(i)a\tilde{T}b})dv_{\tilde{g}}.
\end{equation}

\noindent We will say that a term is an ${\mathcal{O}}(\sigma)$ if it can be bounded in absolute value by 
a term of the form $C\sigma^{n}e^{-\sigma}$ for some natural number 
$n$. Let's start by analyzing the first term on the right hand side of equation \ref{Gaa}. Making

\begin{equation}
\hat{\Pi}_{ab}=\Pi_{ab}+\frac{k}{3}({\bf g}_{ab}+T_{a}T_{b}),\ \hat{N}=N-\frac{3}{k^{2}},
\end{equation}

\noindent we get

\begin{equation}
-9\int_{\Sigma}\tilde{N}Q_{abTT}\tilde{\hat{\Pi}}^{ab}dv_{\tilde{g}}-3\tilde{Q}_{i}-9\int_{\Sigma}\tilde{\hat{N}}\tilde{Q}_{i}dv_{\tilde{g}}.
\end{equation}

\noindent Using Lemma \ref{Estimate} and the estimate on $\tilde{\mathcal{E}}_{1}$ above we get the term

\begin{equation}\label{-3}
-3\tilde{Q}_{i}+{\mathcal{O}}(\sigma)\tilde{Q}_{i}.
\end{equation} 

\noindent Now we estimate the second term in equation \ref{Gaa}, and therefore we need estimates of $\tilde{J}$ and $\tilde{J}^{*}$.
We will do the calculations only for $J$, those for $J^{*}$ proceed in exactly the same way. We note first the 
following inductive formula for $J$

\begin{equation}
J({\bf W}_{i})_{abc}=\hat{\Pi}^{de}\bn_{e}{\bf W}_{(i-1)dabc}-\frac{k}{3}{\bf W}_{(i)dabc}T^{d}+T*{\bf Rm}*{\bf W}_{i-1}
+\bn_{T}J({\bf W}_{i})_{abc}
\end{equation}

\noindent where the $*$ is some tensorial multiplication whose particular form is not important to our purposes. We can write
the formula above symbolically as

\begin{equation}
J({\bf W}_{i})=\hat{\Pi}*\bn {\bf W}_{i-1}-\frac{k}{3}{\bf W}_{i}*T-\frac{k}{3}J({\bf W}_{i-1})+T*{\bf Rm}*{\bf W}_{i-1}
+\bn_{T} J({\bf W}_{i-1}).
\end{equation}

\noindent Now, inducting the fifth term on the first, second, third and fourth gives the following terms respectively

\begin{enumerate}

\item 
\begin{equation}
\sum_{j=0}^{j=i-1}\bn_{T}^{j}(\hat{\Pi}*\bn {\bf W}_{i-1-j})
\end{equation}

\item \begin{equation}\label{ind2}
\sum_{j=0}^{j=i-1}\bn_{T}^{j}(\frac{-k}{3}*T*{\bf W}_{i-j}),
\end{equation}

\item \begin{equation}\label{ind3}
\sum_{j=0}^{i-2}\bn_{T}^{j}(\frac{-k}{3}J({\bf W}_{i-(j+1)})),
\end{equation}

\item \begin{equation}\label{ind4}
\sum_{j=0}^{i-1}\bn_{T}^{j}(T*{\bf Rm}*{\bf W}_{i-1-j}).
\end{equation}

\end{enumerate}

\noindent The only terms that are not going to count as 
${\mathcal{O}}(\sigma)$ or ${\mathcal{O}}(\sigma)\tilde{Q}_{i}^{\frac{1}{2}}$ are
those coming from the expression $2$ and when the $\bn_{T}$ derivative applies only to the ${\bf W}_{i-j}$ 
giving

\begin{equation}
\frac{-k}{3}i{\bf W}_{i}*T
\end{equation} 

\noindent When we take into accound this and a similar term aroused from a formula for $J^{*}$
and plug them into equation \ref{Gaa} we get a contribution of the form

\begin{equation}\label{maineq}
-2i\tilde{Q}_{i}
\end{equation}

\noindent As said above and as we will explain in a moment all other terms are going to count as 
${\mathcal{O}}(\sigma)$ or ${\mathcal{O}}(\sigma)\tilde{Q}_{i}^{\frac{1}{2}}$
therefore we would get, putting equations \ref{31},\ref{-3} and the last estimate together we get

\begin{equation}
\frac{d\tilde{Q}_{i}}{d\sigma}=-(2+{\mathcal{O}}(\sigma))\tilde{Q}_{i}+{\mathcal{O}}(\sigma)\tilde{Q}_{i}^{\frac{1}{2}},
\end{equation}

\noindent as we wanted in the induction. To discuss the other terms then we start by recalling some 
propositions from \cite{R} restated in a different form for convenience of the article.

\begin{Lem}\label{Pi} Let $(g,K)$ be a CMC flow on a rigid hyperbolic manifold $\Sigma$. Suppose that 
the initial cosmological state is $\epsilon$-close to the standard flat cone state $(g_{H},-g_{H})$ as in
Lemma \ref{Estimate} then (all derivatives below are taken at $k=-3$)

\begin{enumerate}

\item \begin{equation}
\|\bn_{T}^{i}\Pi\|_{H^{j}_{g_{H}}},\ i\geq 1,\ j=0,1,2,
\end{equation}

are controlled by ${\mathcal{E}}_{i+j-1}$.

\item \begin{equation}\label{ineq}
(\|\bn {\bf W}_{i})\|_{L^{2}_{g_{H}}}+\|{\bf W}_{i}\|_{L^{2}_{g_{H}}})\leq C(\|{\bf W}_{i+1}\|_{L^{2}_{g_{H}}}+
\|{\bf W}_{i}\|_{L^{2}_{g_{H}}}+\|J({\bf W}_{i})\|_{L^{2}_{g_{H}}})
\end{equation}

$i\geq 0$.

\end{enumerate}
\end{Lem}

\begin{Lem} $\bn_{T}^{h}J({\bf W}_{i})$ has an expression of the form
\begin{equation}\label{J}
\bn_{T}^{h}J({\bf W}_{i}) = \sum (\bn_{T}^{m_{1}} \Pi)^{n_{1}}*\cdots*(\bn_{T}^{m_{s}} \Pi)^{n_{s}}*\Pi^{l}*
                 \bn {\bf W}_{k}+ 
\end{equation}
\begin{equation}\label{J2}
\ \ \ \ \ \ \ \ \ \ \  +\sum (\bn_{T}^{\tilde{m}_{1}} \Pi)^{\tilde{n}_{1}}*\cdots*(\bn_{T}^{\tilde{m}_{s}} 
                   \Pi)^{\tilde{n}_{s}}*\Pi^{\tilde{l}}*\bn^{q}_{T}(T*{\bf Rm}*{\bf W}_{\tilde{k}})                
\end{equation}

\noindent where the first sum is among the set $k\leq i+h-1$, $m_{1}\geq\ldots\geq m_{s}\geq 1$ and 
$\sum_{j}n_{j}(1+m_{j})+l+k=i+h$, while
the second is among the set $\tilde{m}_{1}\geq \ldots \geq \tilde{m}_{s}\geq 1$ and 
$\sum_{j} \tilde{n}_{j}(1+\tilde{m}_{j})+\tilde{k}+\tilde{l}+q=i+h-1$.

\end{Lem} 

\noindent Now we prove the following Lemma.

\begin{Lem}\label{lemf} Let $(g,K)$ be a CMC solution. Suppose for a given value of $i$ there are $n_{i}$ and $C_{i}$ such 
that $\tilde{\mathcal{E}}_{i}\leq C_{i}\sigma^{n_{i}}e^{-2\sigma}={\mathcal{O}}(\sigma)$ then 

\begin{enumerate} \item there are $n'_{i}$ and $C'_{i}$ such that 
$\|\tilde{J}({\bf W}_{i})\|^{2}_{L^{2}_{g_{H}}}\leq C'_{i}\sigma^{n'_{i}}e^{-2\sigma}={\mathcal{O}}(\sigma)$.

\item There are $n'_{ij}$ and $C'_{ij}$ such that 
$\|(\bn_{T}^{j}J({\bf W}_{i-j}))^{\sim}\|^{2}_{L^{2}_{g_{H}}}\leq C'_{ij}\sigma^{n'_{ij}}e^{-2\sigma}={\mathcal{O}}(\sigma)$ 
for $j\leq i$. 

\end{enumerate}

\end{Lem}

\noindent {\bf Proof:} 1. Proceed by induction in $i$. Observe that all the factors involving $\Pi$ and its
time derivatives in formula \ref{J} (with $h=0$) are controlled by $\tilde{\mathcal{E}}_{i}$ in $H^{2}_{g_{H}}$ by
Lemma \ref{Pi}. 
The norms $\|(\bn {\bf W}_{k})^{\sim}\|_{L^{2}_{g_{H}}}$
are controlled using inequality \ref{ineq}. The second kind of terms in equation \ref{J2} are controlled as follows.
The factors involving $\Pi$ and its time derivatives are controlled again in $H^{2}_{g_{H}}$ by 
$\tilde{\mathcal{E}}_{i}$. The other factors can be seen as

\begin{equation}
(\bn^{q}_{T}(T*{\bf Rm}*{\bf W}_{\tilde{k}}))^{\sim}=\sum_{q_{1}+q_{2}+q_{3}=q} (\bn^{q_{1}}_{T}T)^{\sim}*(\bn^{q_{2}}_{T}{\bf Rm})^{\sim}*(\bn^{q_{3}}_{T}{\bf W}_{\tilde{k}})^{\sim},
\end{equation}

\begin{equation}
\ \ \ \ \ \ \ \ \ \ \ =\sum_{q_{1}+q_{2}+q_{3}=q} (\bn^{q_{1}}_{T}T)^{\sim}*(\tilde{\bf W}_{q_{2}})*
(\tilde{\bf W}_{q_{3}+\tilde{k}}),
\end{equation}

\noindent with $q\leq i-1$. Now Sobolev embeddings give

\begin{equation}
\|\tilde{\bf W}_{q_{2}}*\tilde{\bf W}_{\tilde{k}+q_{3}}\|_{L^{2}_{g_{H}}}\leq 
C(\|\tilde{\bf W}_{q_{3}}\|_{H^{1}_{g_{H}}}\|\tilde{\bf W}_{\tilde{k}+q_{3}}\|_{H^{1}_{g_{H}}}),
\end{equation}

\noindent where the factors on the right are controlled by Lemma \ref{Pi}. The factors $(\bn_{T}T)^{\sim}$ are
controlled in $H^{2}_{g_{H}}$ by Lemma \ref{Pi}. Finally the proof of part 2. is the 
same as above after using formulas \ref{J}, \ref{J2}.\hspace{\stretch{1}}$\Box$\\   

The terms in $2,\ 3,\ 4$ on the induction formula for $J$ other than the ones already considered in
equation \ref{maineq} are easily seen to be bounded by ${\mathcal{O}}(\sigma)$ 
or ${\mathcal{O}}(\sigma)\tilde{Q}_{i}^{\frac{1}{2}}$ by the same kind of arguments as in Lemma \ref{lemf}.
To bound the terms in 1. in the same way we need the following form of $\bn_{T}^{j}\bn {\bf W}_{k}$ 

\begin{equation}
\bn_{T}^{j} \bn {\bf W}_{i}=\sum (\bn_{T}^{m_{1}} \Pi)^{n_{1}}*\cdots*(\bn_{T}^{m_{s}} \Pi)^{n_{s}}*\Pi^{l}*
                 \bn {\bf W}_{k}+ 
\end{equation}
\begin{equation}\label{J3}
\ \ \ \ \ \ \ \ \ \ \  +\sum (\bn_{T}^{\tilde{m}_{1}} \Pi)^{\tilde{n}_{1}}*\cdots*(\bn_{T}^{\tilde{m}_{s}} 
                   \Pi)^{\tilde{n}_{s}}*\Pi^{\tilde{l}}*\bn^{q}_{T}(T*{\bf Rm}*{\bf W}_{\tilde{k}}),                
\end{equation}

\noindent where the first sum is among the set $m_{1}\geq\ldots\geq m_{s}\geq 1$ and 
$\sum_{j}n_{j}(1+m_{j})+l+k=i+j$, while
the second is among the set $\tilde{m}_{1}\geq \ldots \geq \tilde{m}_{s}\geq 1$ and 
$\sum_{j} \tilde{n}_{j}(1+\tilde{m}_{j})+\tilde{k}+\tilde{l}+q=i+j-1$, which can be easily proved by induction
by using equation

\begin{equation}
\bn_{T}\bn {\bf W}_{i}=\bn {\bf W}_{i+1}+\Pi*\bn {\bf W}_{i}+ T*{\bf Rm}*{\bf W}_{i}.
\end{equation}
 
\noindent This finishes the induction in Lemma \ref{induction}.\hspace{\stretch{1}}$\Box$.

\noindent {\it Items $3$ and $4$.} The estimate from above in item $4$ comes from Lemma \ref{Estimate}. The
item $3$ or the estimate from below is more involved, the argument is as follows.

\begin{Lem}\label{VvK} For any $\epsilon>0$ there is a ball $B_{(g_{H},-g_{H})}(\delta)$ of cosmologically scaled states 
in $H^{3}\times H^{2}$ such that

\begin{equation}
\|\tilde{N}-\frac{1}{3}\|_{L^{\infty}}\leq \epsilon,
\end{equation}

\noindent and 

\begin{equation}
4\pi G {\mathcal{H}} E_{CMC}\geq \frac{1}{4+\epsilon}\int_{\Sigma}|\hat{\tilde{K}}|^{2}dv_{\tilde{g}}.
\end{equation}

\end{Lem}

We can prove item $3$ by making use of the Lemma 
\ref{VvK}. First, the derivative of the reduced volume ${\mathcal{V}}={\mathcal{H}}^{3}V$ in logarithmic time is 

\begin{equation}
\frac{d{\mathcal{V}}}{d\sigma}=-3\int_{\Sigma}\tilde{N}|\hat{\tilde{K}}|^{2}dv_{\tilde{g}}.
\end{equation}

\noindent If we integrate it from $\sigma$ to $\infty$ and use lemma \ref{VvK} above we get the following inequality

\begin{equation}\label{ineq}
\frac{1}{4+\epsilon}\int_{\Sigma}|\hat{\tilde{K}}|^{2}dv_{\tilde{g}}\leq 4\pi G {\mathcal{H}} E_{CMC}=
3\int_{\sigma}^{\infty}(\int_{\Sigma}\tilde{N}|\hat{\tilde{K}}|^{2}dv_{\tilde{g}})
d\sigma\leq (1+\epsilon)\int_{\sigma}^{\infty}(\int_{\Sigma}|\hat{\tilde{K}}|^{2}dv_{\tilde{g}})d\sigma.
\end{equation}
 
\noindent Making $U=\int_{\sigma}^{\infty}(\int_{\Sigma}|\hat{\tilde{K}}|^{2}dv_{\tilde{g}})d\sigma$ the inequality
\ref{ineq} is written as

\begin{equation}
U'\geq -(4+\epsilon)(1+\epsilon)U,
\end{equation}

\noindent which after integration gives the left hand side inequality in item $3$.

\noindent {\bf Proof of Lemma \ref{VvK}}. First we note that the estimate for $\tilde{N}-\frac{1}{3}$ is deduced
from Lemma \ref{Estimate}. For the second estimate it may be deduced from the calculation of the Hessian of the
energy that we did before, however we will follow a direct estimate from the Lichnerowicz equation. We argue 
as follows. Say $g=\phi^{4}g_{Y}$ where $g_{Y}$ is the unique metric in the conformal class of $g$ having
scalar curvature $-6$. Then $\phi$ satisfies

\begin{equation}\label{Lichn}
-\Delta \phi +\frac{3}{4}(\phi^{5}-\phi)=\frac{1}{8}\phi^{-3}|\hat{\tilde{K}}|^{2}_{Y}.
\end{equation}

\noindent The maximum principle gives $\phi\geq 1$. Making $\bar{\phi}=\phi-1$ rewrite equation \ref{Lichn} 
as 

\begin{equation}
-\Delta \bar{\phi}+\frac{3}{4}\phi(\phi^{3}+\phi^{2}+\phi+1)\bar{\phi}=\frac{|\hat{\tilde{K}}|^{2}_{Y}}{8\phi^{3}}.
\end{equation}

\noindent At the point where $\phi$ or $\bar{\phi}$ is maximum we have

\begin{equation}
\bar{\phi}\leq \frac{1}{12}\frac{|\hat{\tilde{K}}|^{2}\phi^{4}}{\phi^{3}+\phi^{2}+\phi+1}\leq \frac{|\hat{\tilde{K}}|^{2}\phi}{12},
\end{equation}

\noindent which gives if $\|\hat{K}\|_{L^{\infty}_{g}}$ is small

\begin{equation}\label{linf}
\|\bar{\phi}\|_{L^{\infty}}\leq \frac{\|\hat{K}\|^{2}_{L^{\infty}_{g}}}{12-\|\hat{K}\|^{2}_{L^{\infty}_{g}}}.
\end{equation}

\noindent Also note that 

\begin{equation}
-\sigma(\Sigma)\leq -6V_{Y}^{\frac{2}{3}},
\end{equation}

\noindent which gives

\begin{equation}
0\leq \int_{\Sigma}(\phi^{6}-1)dv_{Y}\leq V-V_{H}.
\end{equation}

\noindent Writing $\phi^{6}-1=(\phi-1)(\phi^{5}+\phi^{4}+\phi^{3}+\phi^{2}+\phi+1)$ we get

\begin{equation}
6\int_{\Sigma}(\phi-1)dv_{g_{Y}}\leq V-V_{H}.
\end{equation}

\noindent Integrating equation \ref{Lichn} we get 

\begin{equation}\label{abo}
6\int_{\Sigma}(\phi^{5}-\phi)dv_{g_{Y}}=\int_{\Sigma}\phi^{-3}|\hat{\tilde{K}}|^{2}_{Y}dv_{g_{Y}}.
\end{equation}

\noindent Under the assumptions we have and using equation \ref{linf} we can get from equation \ref{abo} above
the inequality

\begin{equation}
6(4+\epsilon)\int_{\Sigma}(\phi-1)dv_{g_{Y}}\geq \int_{\Sigma}\phi^{-2}|\hat{\tilde{K}}|^{2}_{Y}dv_{g_{Y}}=\int_{\Sigma}|\hat{\tilde{K}}|^{2}dv_{g}.
\end{equation} 

\noindent which together with equation \ref{final} gives the inequality

\begin{equation}
(4+\epsilon)(V-V_{H})\geq \int_{\Sigma}|\hat{\tilde{K}}|^{2}dv_{g}.
\end{equation}

\noindent as desired.\hspace{\stretch{1}}$\Box$.

This finishes theorem \ref{Smooth}.\hspace{\stretch{1}}$\Box$.

{\center \section{States of arbitrarily large gravitational energy.}\label{sec5}}

\vspace{0.1cm} 
We will construct a one parameter family of states $(g_{\lambda},K_{\lambda})$ such that\\

\begin{enumerate}
\item $k_{\lambda}=k_{0}$ fixed,
\item $Vol_{g_{\lambda}}\rightarrow_{\lambda\rightarrow \infty} \infty$ and $\|\hat{K}_{\lambda}\|_{L^{2}_{g_{\lambda}}}\rightarrow_{
\lambda\rightarrow \infty} 
\infty$,
\item The ``big-bang" family of states, i.e. the volume-one normalized family of states above has 

\begin{equation}
-k_{\lambda}\rightarrow \infty,
\end{equation}

\begin{equation}
Vol_{g_{\lambda}}(\Sigma)=1,
\end{equation}

\begin{equation}
\lim_{\lambda\rightarrow \infty}\|\hat{K_{\lambda}}\|_{L^{2}_{g_{\lambda}}}=\infty.
\end{equation}

\end{enumerate} 

\noindent As has been argued above, these states represents a one parameter family of states with arbitrarily large
gravitational energy. The construction is as follows. Pick the hyperbolic metric $g_{H}$ and a non zero transverse
traceless tensor $\hat{K}$ with respect to it. According to the conformal method it is possible to find a solution to the
constraint of the form $(g_{\lambda},K_{\lambda})=(\varphi^{4}g_{H},\lambda^{2}\varphi^{-2}\hat{K}-\varphi^{4}g_{H})$
(the mean curvature being $k=k_{0}=-3$ and one parameter family of states as above with arbitrary $k_{0}$ can
be obtained by scaling), by solving the elliptic equation

\begin{equation}\label{Y}
\Delta \varphi=-\frac{3}{4}\varphi -\frac{\lambda^{4}}{8}|\hat{K}|^{2}_{g_{H}}\varphi^{-7}+\frac{3}{4}\varphi^{5}.
\end{equation}

\noindent Now we prove items $2$. Multiplying equation \ref{Y} by $\varphi$ and integrating we
get

\begin{equation}\label{Yint}
\frac{\lambda^{4}}{8}\int_{\Sigma}|\hat{K}|_{g_{H}}^{2}\varphi^{-6}dv_{g_{H}}=\int_{\Sigma}|\nabla \varphi|^{2}+\frac{3}{4}(\varphi^{6}-\varphi^{2})dv_{g_{H}}.
\end{equation}

\noindent Note that the left hand side is $\frac{1}{8}\|\hat{K}_{\lambda}\|^{2}_{L^{2}_{g_{\lambda}}}$. If the left hand
side doesn't diverge as $\lambda\rightarrow \infty$ then the right hand side remains bounded in particular 
the $H^{1}_{g_{H}}$ norm of $\varphi$ remains bounded. Pick an open set $\Omega$ where $|\hat{K}|_{g_{H}}\geq \epsilon>0$.
Then as $\varphi$ is bounded in $H^{1}$ we have $Vol\{x\in \Omega/\varphi(x)<n\}\rightarrow Vol(\Omega)$ as $n\rightarrow \infty$
uniformly in $\lambda$. Then for some $n$ we have $Vol\{x\in \Omega/\varphi(x)<n\}> \frac{Vol(\Omega)}{2}$ uniformly in $\lambda$,
and so the left hand side is bigger than $\frac{\lambda^{4}}{16 n^{6}}\epsilon^{2}Vol(\Omega)$ which diverges when
$\lambda\rightarrow \infty$ which is a contradiction. This proves item $2$, to prove item $3$ we argue as follows. 
The $L^{2}$ norm of $\hat{K}_{\lambda}$ of the volume one states are

\begin{equation}
\frac{\lambda^{4}\int_{\Sigma}|\hat{K}|^{2}\varphi^{-6}dv_{g_{H}}}{(\int_{\Sigma}\varphi^{6}dv_{g_{H}})^{\frac{1}{3}}}
=\frac{\int_{\Sigma}|\nabla \varphi|^{2}+\frac{3}{4}(\varphi^{6}-\varphi^{2})dv_{g_{H}}.}
{(\int_{\Sigma}\varphi^{6}dv_{g_{H}})^{\frac{1}{3}}}.
\end{equation}

\noindent We have that an upper bound on the left hand side in the last equation implies an upper bound for
the $H^{1}$ norm of $\varphi$, for if not we have $\int_{\Sigma}\varphi^{6}dv_{g_{H}}\rightarrow \infty$ which
would make the numerator of the right hand side diverging in $\lambda$, but we know $\int_{\Sigma}\varphi^{6}dv_{g_{H}}$
diverges which is a contradiction.

{\center \section{Summary and open questions.}\label{sec7}}

\vspace{0.1cm}
We have introduced the notion of general ${\mathcal{K}}=-1$ cosmological model as a formal definition allowing to 
study cosmological notions in arbitrary solutions of the Einstein equations. This gave us a framework
to study general cosmological solutions in a cosmological language. The approach may be applicable to models other
than general ${\mathcal{K}}=-1$ cosmological models, i.e. models with different spatial topologies. Thinking on
the {\it averaging problem in cosmology} we have defined volume-averaged cosmological parameters and an averaging
map: a correspondence between arbitrary solutions and homogeneous and isotropic Lorentzian spaces. Those concepts 
allowed us to give a precise mathematical formulation
of the averaging problem in cosmology. In another section and aiming at the start of a rigorous analysis of cosmological 
evolution from the solutions at the natural scale, (i.e. including the small scale), we have introduced {\it assumption (C)} 
which precisely describe a certain class of solutions. Those solutions are divided into two main subclasses: radiative and mass gap. We have 
given a detailed description of the full structure of the radiative 
solutions. We have also analyzed the averaging problem in cosmology in precise quantitative terms for mass gap 
solutions. The attemp may be considered as a first step towards the ideal goal of attacking the averaging problem in cosmology 
directly from the solutions at the small scale. Finally
we constructed initial ``big-bang" states of arbitrarily large gravitational energy, showing that, apriori there is no mathematical
restriction to assume the gravitational energy to be low at the beginnings of time.

There are several questions and avenues of research left open in the present article, of varying difficulty however. 
For instance one may want to 
see in action the formalism of {\it general cosmological models} in cosmological solutions with Cauchy surfaces of
non hyperbolic topology. Also and perhaps more important is to obtain rigorous results that may support or not 
{\it assumption (C)}. Any rigorous result of the sort would put the study of the averaging problem in cosmology 
from the small 
scale on a firm basis. Analyzing the validity of assumption C from the Einstein equations is a very 
difficult problem. A central point is to
study the spatial asymptotic of stationary solutions that may emerge in time. Is an emerging stationary solution 
necessarily spatially asymptotically flat in the long time?. If the answer is affirmative one may be in a better position to 
prove the a priori estimates in assumption (C). The answer may instead be negative and that would open a new 
avenue of research. Finally the analysis of the validity of the averaging problem in cosmology from assumption C was 
only asymptotic in time, and therefore of non obvious applicability. An interesting question is to study the 
validity of the analysis but in finite times.

\end{document}